\documentclass[11pt,letterpaper]{article}

\usepackage{amsfonts,amssymb,amsmath,amsthm,url}

\usepackage[boxed]{algorithm2e} 
\dontprintsemicolon

\usepackage{graphics}

\usepackage{color}
\usepackage[all]{xy}
\SelectTips{cm}{}

\newcommand{\br}[1]{\left( #1 \right)}

\newcommand{\bc}[1]{\left\{ #1 \right\}}

\newcommand{\bg}[1]{\left< #1 \right>}

\newcommand{\imps}{\ \ \Longrightarrow\ \ }
\newcommand{\cimps}{\Longrightarrow}

\newcommand{\twodefo}[3]{\ensuremath{\left\{\begin{array}{ll}#1&\mbox{if\ \ }#2\\#3&\mbox{otherwise}\end{array}\right.}}
\newcommand{\otwodef}[4]{\ensuremath{\left\{\begin{array}{ll}#1&\mbox{if\ \ }#2\\#3&\mbox{otherwise if }#4\end{array}\right.}}

\newcommand{\threedefo}[5]{\ensuremath{\left\{\begin{array}{ll}#1&\mbox{if\ \ }#2\\#3&\mbox{if\ \ }#4\\#5&\mbox{otherwise}\end{array}\right.}}

\newcommand{\fmod}[1]{\ensuremath{\ \mbox{mod } #1}}

\newcommand{\twostack}[2]{\ensuremath{\left\{\begin{array}{l}#1\\#2\end{array}\right.}}

\newcommand{\tab}{\hspace{1cm}}

\newcommand{\goesto}{\rightarrow}

\newcommand{\wild}{-}

\newcommand{\Z}{\ensuremath{\mathbb{Z}}}

\renewcommand{\O}{\ensuremath{{\cal O}}}

\newcommand{\algreturn}{\textbf{return} \ }

\newcommand{\remove}[1]{}

\newcommand{\we}{\ensuremath{\,\wedge\,}}
\newcommand{\ve}{\ensuremath{\,\vee\,}}

\renewcommand{\L}[1]{{\bar{#1}}}

\newcommand{\MQ}{MutexQueue}
\newcommand{\Wayte}{\mathit{Wait}}
\newcommand{\rt}{\mathit{ret}}

\newcommand{\enqueue}{\ensuremath{\FuncSty{enqueue()}}}
\newcommand{\isHead}{\ensuremath{\FuncSty{isHead()}}}
\newcommand{\dequeue}{\ensuremath{\FuncSty{dequeue()}}}
\newcommand{\opread}{\ensuremath{\FuncSty{read}}}
\newcommand{\opwrite}{\ensuremath{\FuncSty{write}}}

\newcommand{\OK}{\ensuremath{\FuncSty{OK}}}
\newcommand{\INV}{\ensuremath{\FuncSty{INV}}}
\newcommand{\RES}{\ensuremath{\FuncSty{RES}}}
\newcommand{\ATM}{\ensuremath{\FuncSty{ATOM}}}
\newcommand{\FI}{\ensuremath{\FuncSty{F\&I}}}
\newcommand{\Sw}{\ensuremath{\FuncSty{F\&S}}}

\newcommand{\true}{\ensuremath{\FuncSty{true}}}
\newcommand{\false}{\ensuremath{\FuncSty{false}}}

\newtheoremstyle{coolstyle}
    {16pt}
    {9pt}
    {\slshape}
    {}
    {\bfseries}
    {.}
    {.5em} 
    {}

\theoremstyle{coolstyle}
\newtheorem{theorem}{Theorem}[section]
\newtheorem{observation}[theorem]{Observation}

\newtheorem{lemma}[theorem]{Lemma}
\newtheorem{invariant}[theorem]{Invariant}
\newtheorem{corollary}[theorem]{Corollary}

\remove{

\SetKw{Wait}{wait until}

\SetKwFunction{ID}{PID}

\SetKwFunction{ForestCC}{Forest-CC}
}

\begin{document}

\title{Deconstructing Queue-Based Mutual Exclusion}
\author{Wojciech Golab\thanks{Research conducted during doctoral studies at the University of Toronto, Canada.}\,\,\thanks{
Research partially supported by the Natural Sciences and Engineering Research Council (NSERC) of Canada.}\\
\url{wgolab@uwaterloo.ca} \\ \ \\
Hewlett-Packard Labs Technical Report HPL-2012-100 \\
\date{May 6, 2012}
}

\maketitle

\begin{abstract}
We formulate a modular approach to the design and analysis of
a particular class of mutual exclusion algorithms for
shared memory multiprocessor systems.
Specifically, we consider algorithms that organize waiting processes into a queue.
Such algorithms can achieve $\O(1)$ remote memory reference (RMR) complexity,
   which minimizes (asymptotically) the amount of traffic through the processor-memory interconnect.
We first describe a generic mutual exclusion algorithm that relies
on a linearizable implementation of a particular queue-like data structure
that we call \emph{\MQ}.
Next, we show two implementations of \MQ\ using $\O(1)$ RMRs per operation
based on synchronization primitives commonly available in multiprocessors.
These implementations follow closely the queuing code embedded in
previously published mutual exclusion algorithms.
We provide rigorous correctness proofs and RMR complexity analyses of the
algorithms we present.
\end{abstract}



\section{Introduction \label{sec_intro}}
Synchronization is a fundamental challenge in asynchronous
shared memory multiprocessor systems,
where processes executing in parallel must exercise caution while accessing shared data structures.
Unless concurrent access to such data structures is directly supported in hardware,
careful coordination is necessary at the software level
to prevent corruption of the data structure and ensure that processes executing
on different processors reach consistent views of the data.
The dominant approaches to such coordination are mutual exclusion,
and non-blocking synchronization.

Mutual exclusion (ME) was formulated by Dijkstra \cite{dijk:soln},
and later formalized by Lamport \cite{lamp:par1, lamp:par2}.
In this approach, processes take turns accessing the shared data structure.
The execution path of each process is modelled as a repeating sequence of
four sections, illustrated below in Figure~\ref{fig_mutex}.
Access to the shared data structure is confined to a
special \emph{critical section} (CS), to which the process must gain
exclusive access by executing the \emph{entry protocol}.
Similarly, an \emph{exit protocol} is executed upon completing the CS
to signal that the CS can now be entered by another process.
Between executions of the CS, a process lives in the \emph{non-critical section} (NCS).
The set of variables accessed by a process while in the CS or the NCS is disjoint from the
set of variables accessed while in the entry or exit protocol.

An execution by a process of the entry protocol, CS and exit protocol is referred
to as a \emph{passage} (through the ME algorithm).
The entry section is sometimes divided into a \emph{doorway}, where a process enters
a queue by executing a bounded number of steps, and a \emph{waiting room}, where
it waits for its predecessor in the queue to exit the CS.
This type of algorithm is referred to as \emph{first-come first-served} (FCFS).
\emph{Lockout freedom} (also referred to as \emph{starvation freedom}) is a progress
 whereby every process
that begins the entry protocol eventually enters the CS, provided that no process
halts outside the NCS.
Finally, \emph{deadlock freedom} is a weaker property that
guarantees the same but only provided that no process passes through the CS
infinitely often, and is considered the weakest progress property required
for correctness of a mutual exclusion algorithm.

    \begin{figure}
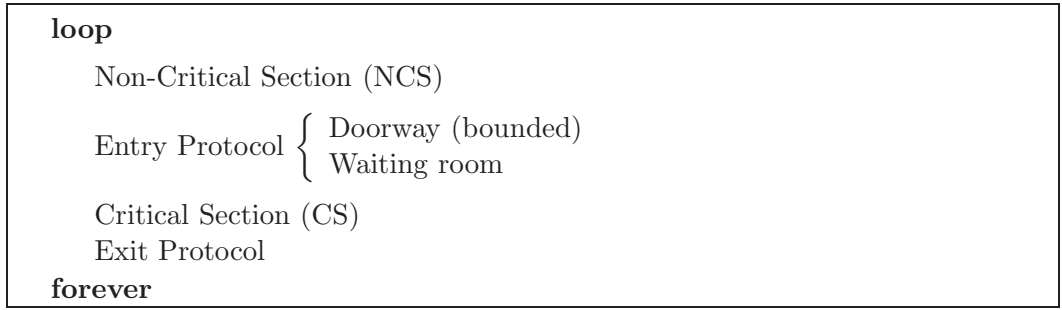

    \begin{center}
  \begin{algorithm}[H]
  \SetKwBlock{LoopForever}{loop}{forever}
  \LoopForever{
    \vspace{6pt}
    Non-Critical Section (NCS) \;
    \vspace{6pt}
    Entry Protocol \twostack{\mbox{Doorway (bounded)}}{\mbox{Waiting room}}  \; 
    \vspace{6pt}
    Critical Section (CS)  \;
    Exit Protocol   \;
    \vspace{6pt}
    }
	\end{algorithm}
    \end{center}
    \caption{Execution path of a process participating in mutual exclusion. \label{fig_mutex}}
    \end{figure}

In contrast to mutual exclusion, non-blocking synchronization requires some measure
  of progress regardless of the rates at which other processes are executing.
For example in wait-free synchronization \cite{herl:wait}, a process
  must complete each access to the shared data structure in a finite number of their own steps.
The key idea behind universal constructions of wait-free data structures
  is that faster processes assist slower ones in performing updates.
To that end, processes exploit hardware synchronization primitives to agree
  on the order in which updates are applied, and hence on the state of the shared data.

Mutual exclusion and wait-freedom have complementary characteristics.
ME is a \emph{blocking} approach since a fast process can spend an unbounded
amount of time in a busy-wait loop, which typically involves repeatedly
testing or \emph{spinning on} one variable, while waiting for another process
to complete the CS and then write that variable.
In contrast, in a wait-free algorithm a process
ensures its own progress even if all others halt at an arbitrary point
in their execution.
In that sense, wait-freedom is a stronger progress property than the one
  underlying mutual exclusion.
Not surprisingly, there exist shared object types for which 
wait-free implementations are provably more costly than blocking ones.
For example, any wait-free $N$-process implementation of fetch-and-increment using
atomic read/write registers (subsequently
referred to simply as registers) and fetch-and-store requires $\Omega(\log N)$
remote memory references (RMR, discussed below) \cite{jaya:lower}.
In contrast, there is a blocking implementation of fetch-and-increment 
from registers and fetch-and-store using only $\O(1)$ RMRs.
In light of its lower cost, mutual exclusion remains the dominant approach
  to synchronization in practice.

\remove{
    \begin{table}
    \begin{center}
    \begin{tabular}{|c|c|c|} \hline\hline
    Mechanism & Blocking? & Complexity \\  \hline \hline
    wait-freedom & no &  higher \\
    mutual exclusion & yes & lower \\   \hline\hline
    \end{tabular}
    \end{center}
    \caption{Comparison of wait-freedom and mutual exclusion. \label{fig_compa}}
    \end{table}
}

\subsection{Time Complexity of Mutual Exclusion Algorithms}
\begin{figure}
    \begin{center}
    \scalebox{0.75}{\includegraphics*[0cm, 12cm][16cm,23cm]{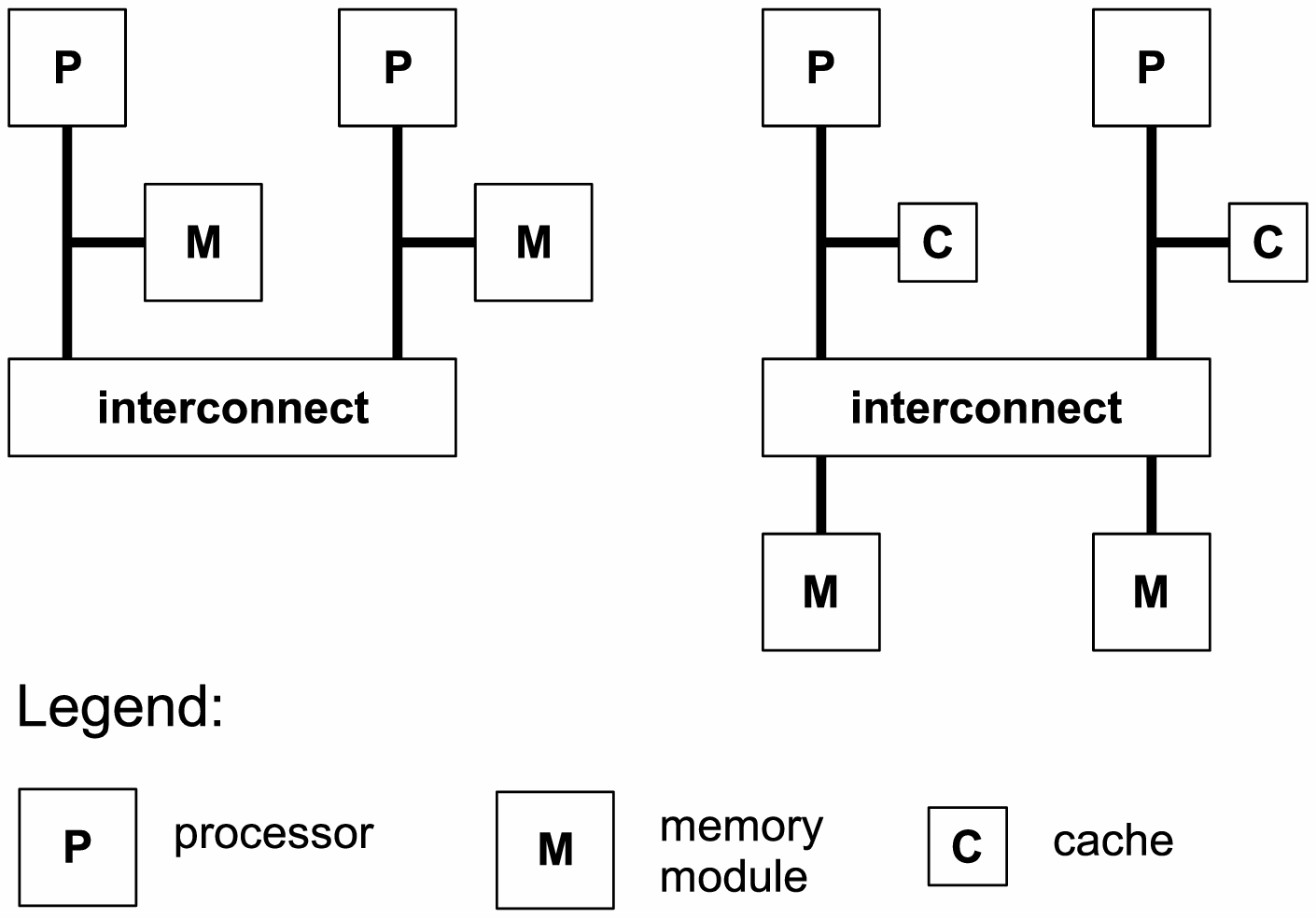}}
    \end{center}
    \caption{Shared memory architectures: DSM (left) and CC (right).\label{fig_arch}}
    \end{figure}

Analyzing the time complexity of ME algorithms requires some awareness
of the shared memory hardware architecture as different memory operations
may incur significantly different latencies.
This is due to the growing disparity between processor speed and memory 
access speed, which motivates multiprocessor designs based on the 
paradigms of non-uniform memory access (NUMA) and/or caching.
Two important classes of such architectures are illustrated in Figure~\ref{fig_arch}
\cite{mcs:algo, jand:surv}. In a Distributed Shared Memory (DSM) machine, each memory
module can be accessed locally by some processor without involving the
processor-to-memory interconnect, thus reducing much of the latency.
Processors in cache-coherent (CC) machines, on the other hand, maintain
local copies of data inside caches, which are synchronized by a
coherence protocol.  Thus, any shared memory location can become local
at runtime to any processor in the CC model.

In both the DSM and CC models, memory operations are classified 
as either remote or local.
This classification is straightforward in the DSM model as locality is determined
through static allocation of a variable in particular a memory module.
In contrast, in the CC model locality of a memory operation is determined
  by the state of the processor's cache, which depends on prior
  steps of the same processor and possibly others,
  as well as by the type of memory access (e.g., read versus write),
  which determines the behaviour of the coherence protocol.
For our purposes, we consider the following ideal behaviour in the CC model:
  after a processor reads a variable, this variable is held in that
  processor's cache and can be read locally (i.e., without incurring an RMR)
  until another processor writes the same variable.
In both the DSM and CC models, we will assume for worst-case analysis
  that each process runs on a distinct processor.

Remote operations, referred to as \emph{remote memory references} or RMRs,
can be orders of magnitude more costly than local ones.
Consequently, RMR complexity quantifies not only the overhead of accessing 
the processor-to-memory interconnect, but also the main source of latency
incurred while executing a mutual exclusion algorithm.
Mutual exclusion algorithms with bounded RMR complexity are referred to as 
\emph{local-spin}, and have been the focus of recent research \cite{jand:surv}
on shared memory multiprocessors.
In such an algorithm, all busy-waiting must be done by
spinning on locally accessible variables.

To obtain a more direct measure of time complexity, one can consider
the overhead of contention (for the processor-to-memory interconnect and
shared memory modules) in addition to RMRs.
This overhead can be quantified by counting \emph{memory stalls} under the assumption
that concurrent accesses to a common shared variable are serialized \cite{dwork:cont}.
In that case, the $i$'th process in the serialization order incurs
$i-1$ memory stalls as it waits for its predecessors to complete their operations.
The exact overhead of contention depends on the shared memory architecture.
Most notably, in a bus-based system a snooping protocol makes it possible for
multiple processes to read a common shared variable simultaneously.
In that case one counts memory stalls for concurrent writes but not for concurrent reads.

Time complexity measures for mutual exclusion algorithms typically omit
local memory operations.  Although local operations do have an impact on the
overall latency, such a complexity measure is generally unbounded.
Even if the CS is empty, due to the assumption of asynchrony there is no bound on
the time that a process leaving the CS takes to execute the exit protocol
and allow the next process to proceed into the CS.
Furthermore, in an algorithm using only registers, even the first process
to enter the CS may perform an unbounded number of steps unless it is executing
solo \cite{alur:res}.
It is possible to circumvent this problem by defining time complexity in terms
of a virtual clock that ticks once for every interval of time in which 
every process has been given sufficient time to perform one operation
on a shared memory object.
The \emph{response time} of a mutual exclusion algorithm is the number of such
clocks ticks from the time a process leaves the NCS to the time it enters the CS \cite{choy:adaptive}.

\subsection{Contributions of This Paper} \label{sec_contribs}
\remove{
In this paper, we analyze local-spin mutual exclusion algorithms,
based on \cite{tand:spin} and \cite{craig:queue}, that attain $\O(1)$ worst case RMR
complexity in both the CC and DSM models.
We separate the simple high-level idea on which these algorithms are based, namely,
a queue that determines the order in which processes gain access to the CS, from the
somewhat messy details of how the queue is implemented given appropriate hardware support.
We believe that this separation of concerns results in easier-to-follow and more elegant
correctness proofs.  Because of its modularity it potentially simplifies the design of
other algorithms that are based on the same high-level idea but differ on the 
implementation of the process queue (which perhaps uses different primitives).
}

Consider the following simple and intuitively appealing idea
	for an FCFS mutual exclusion algorithm:
Processes wait in a queue to enter the CS.
Only the head of the queue may enter the CS.
A process leaving the NCS adds itself to the end of the queue and,
	if it is not the head of the queue,
	it waits by repeatedly reading a local spin variable.
A process leaving the CS
	removes itself from the (head of the) queue.
    It then writes a shared variable to signal its successor (now the new head of the queue),
    perhaps after checking if such a process exists, to stop waiting and proceed into the CS.
    \remove{\footnote{
    Note to Vassos: the Graunke-Thakkar algorithm does not check if a successor exists, and
                    it writes a spin variable that (re)used by many processes rather than
                    belonging to the successor (since the algo. operates in the CC model and not DSM).
    }}

Clearly, race conditions can arise
	when a process contending for entry to the CS
	checks whether it is the head of the queue
	(perhaps as it does so another process is about to enter the queue),
	and when a process leaving the CS checks whether
	there is a successor in the queue
	(perhaps as it does so another process is about to become
	its successor).
Handling these race conditions properly,
	while relying on standard synchronization primitives and
	using as few RMRs as possible, is a delicate task.

Several algorithms based on the above idea have appeared
	in the literature \cite{tand:spin, graunke:synch, mcs:algo, craig:queue, mag:queue, rhee:fifo, hh:qalg}.
The common simple structure underlying all these algorithms, however,
	is obscured by the intricate details of handling
	the race conditions described above.
Furthermore, to our knowledge, some of these algorithms have not
	been proved correct.

In this paper we propose a modular approach to the design and
	analysis of such algorithms.
We first define a queue-like shared data structure,
	called \emph{\MQ}.
This data structure allows a process to add itself to the end of the queue,
	query whether it is the head of the queue,
	and remove itself from the head of the queue
	(simultaneously determining the identity of its successor
	in the queue, if one exists).
We then present a very simple
	generic queue-based mutual exclusion algorithm
	along the lines described above,
	that uses this data structure as a ``black box''.
We prove the correctness of this algorithm based on the
	abstract properties of \MQ.
This algorithm uses only a constant number of RMRs,
	beyond what are needed to implement the ``black box'' \MQ,
	and applies only a constant number of operations on \MQ,
	per passage.

We then present two implementations of \MQ,
	both using only a constant number of RMRs for each operation
	in the DSM and CC models.
The first uses registers and the fetch-and-increment primitive
    (which atomically increments a shared memory word and returns its previous value)
	while the second uses registers and the fetch-and-store primitive
    (which atomically assigns a new value to a shared memory word and returns its previous value).

The two implementations of \MQ\ are not novel:
	they are embedded in previously published
	mutual exclusion algorithms;
	here, we have simply recast them as implementations
	of the \MQ\ data structure.
Specifically, the first implementation of \MQ\
	is based on a mutual exclusion algorithm due to Tom Anderson~\cite{tand:spin},
	as subsequently modified by James Anderson and Yong-Jik Kim.
The second implementation of \MQ\ is based on
	a mutual exclusion algorithm due to Craig~\cite{craig:queue}.
To our knowledge, however, these mutual exclusion algorithms
	have not been proved correct.\footnote{
A variant of Craig's algorithm \cite{craig:queue} is proved correct in \cite{hh:qalg}.
This variant is intuitively simpler, but uses an array of length $2N$ instead of $N+1$ 
to encode the queue of processes waiting to enter the critical section.}
In this paper we give rigorous correctness proofs of these algorithms
	(as implementations of \MQ).

The advantage of our modular approach is that it
	``factorizes'' the common structure of some queue-based algorithms, 
	in the form of the generic mutual exclusion algorithm.
The correctness of this common part need only be proved once.
What is left in each of these algorithms,
	can be viewed as an implementation of
	the MutexQueue data structure.

Our definition of the MutexQueue also sheds light on how exactly processes coordinate
    access to the critical section in queue-based mutual exclusion algorithms.
For example, in such algorithms a process does not enter the queue
    and also discover whether it became the head element in one atomic step.
Rather, two atomic steps are required, and are therefore represented by distinct MutexQueue operations.
In contrast, a process can exit the queue and discover its successor in one atomic step.
Surprisingly, sometimes a process can also exit the queue and discover
    no other process, even though a successor does exist!
In particular, this occurs if the successor has entered the queue but has not yet
    queried the head element.
Thus, it is the latter step (i.e., querying the head)
    that makes a process ``visible'' to its predecessor,
    and not the mere act of entering the queue.

\section{Related Work \label{sec_relwork}}
The RMR complexity of mutual exclusion algorithms is a function of the number of processes, $N$.
The best known upper bound on the worst-case RMR complexity per passage of algorithms based
on (atomic) read/write registers only is $\O(\log N)$ \cite{yang:fast, kim:lock}.
This bound is tight \cite{atti:rmr}.
The same tight bound holds for the class of mutual exclusion algorithms that
   in addition to registers use \emph{compare-and-swap} (CAS) or
   \emph{load-linked/store-conditional} (LL/SC) -- primitives that conditionally
   change the value of a shared memory location \cite{ghhw:cas}.

Using synchronization primitives such as \emph{fetch-and-store} (i.e., swap between shared memory
and a private register) and \emph{fetch-and-increment},
it is possible to devise mutual exclusion algorithms with worst-case RMR complexity
of only $O(1)$ \cite{tand:spin, graunke:synch, mcs:algo, craig:queue, mag:queue, rhee:fifo, hh:qalg}.\footnote{
    The original algorithm of T.\ Anderson \cite{tand:spin} uses a constant number of RMRs in the CC model
    but is not local-spin in the DSM model.  
    A constant-RMR DSM variant using the same synchronization primitives
    can be obtained by applying the transformation described in \cite{hh:xform}
    or in footnote 7 of \cite{jand:phi2}.
    Rhee's algorithm \cite{rhee:fifo} is targeted at a variant of the DSM model
    with weaker memory consistency,
    where read/write operations executed by one processor may appear to take effect
    in a different order to another processor due to buffering of writes.
    In this model, a special \emph{fence} operation is used to force previously buffered writes to 
    take effect globally.}
The properties of these algorithms are summarized in Table~\ref{tab_algoprop}.
All of these algorithms are based on the concept of a process queue, which determines the order
in which processes enter the CS and enables efficient signaling between processes
that enter the CS consecutively.  
Thus, in addition to mutual exclusion and lockout freedom, these algorithms also
satisfy FCFS.

\begin{table}
\begin{center}
    \begin{tabular}{|c|c|c|c|} \hline
      Publication  & \multicolumn{2}{|c|}{RMR complexity} & Synchronization primitives \\
      reference  & CC model & DSM model & ($+$ read/write registers)  \\ \hline\hline
      \cite{tand:spin} & \O(1) & unbounded & Fetch-and-Increment (unbounded counter) \\ \hline
      \cite{graunke:synch} & \O(1) & unbounded & Fetch-and-Store \\ \hline
      \cite{mcs:algo} & \O(1) & \O(1) & Fetch-and-Store $+$ Compare-and-Swap\\ \hline
      \cite{craig:queue} & \O(1) & \O(1) & Fetch-and-Store \\ \hline
      \cite{mag:queue} & \O(1) & unbounded & Fetch-and-Store $+$ Compare-and-Clear \\ \hline
      \cite{rhee:fifo} & \O(1) & \O(1) & Fetch-and-Store   \\ \hline
      \cite{hh:qalg}   & \O(1) & \O(1) & Fetch-and-Store   \\ \hline
    \end{tabular}
    \caption{Properties of several constant RMR mutual exclusion algorithms. \label{tab_algoprop}}
\end{center}
\end{table}

Many of the queue-based constant-RMR mutual exclusion algorithms cited above were presented
in the context of performance studies, and lack rigorous proofs of correctness.
Moreover, to our knowledge the only
attempt to generalize or unify these algorithms, all of which are based on the process queue
concept, is the generic algorithm of Anderson and Kim \cite{jand:phi2}.
This algorithm solves mutual exclusion using $\O(1)$ RMRs per passage given
a suitable shared-memory primitive fetch-and-$\phi$, which 
   corresponds to the (atomic) execution of the pseudocode shown in Figure~\ref{fig_phi}.

\begin{figure}
    \begin{tabbing}
        \tab\tab \= fetch\_and\_$\phi(var, input)$ \\
        \>1. \tab \= $old := var$ \\
        \>2. \>  $var := \phi(old, input)$ \\
        \>3. \> \algreturn $old$
    \end{tabbing}
   \caption{Fetch-and-$\phi$ primitive. \label{fig_phi}}
\end{figure}

The fetch-and-$\phi$ primitive can be instantiated to a variety
of shared-memory primitives by choosing a suitable function $\phi$.
For example, a fetch-and-store corresponds to
    \[ \phi(old, input) \equiv  input \]
Similarly, if we use $input$ to encode a pair of values $(a, b)$, a compare-and-swap corresponds to
    \[ \phi(old, (a, b)) \equiv \twodefo{b}{old = a}{old} \]
where $a$ and $b$ are the expected and target value of compare-and-swap.
Thus, fetch-and-$\phi$ generalizes various types of read-modify-write primitives,
including conditionals.

Unlike its predecessors, the generic fetch-and-$\phi$ algorithm of \cite{jand:phi2} 
uses two process queues instead of one, in order to cope with the generic and limited
assumptions on the behaviour of the fetch-and-$\phi$ primitive.
Consequently, an additional mechanism is needed to control access to the critical section,
and the algorithm loses the (FCFS) property inherent in earlier single-queue solutions.

Correctness of the generic algorithm depends on a condition on the fetch-and-$\phi$
primitive related to its ability to return distinct values over repeated invocations.
This condition is formalized in terms of a property of a primitive called \emph{rank}.
Intuitively, the higher the rank, the better the primitive at solving
    mutual exclusion efficiently with respect to RMR complexity.
A rank of $2N$ or greater is sufficient for
the generic algorithm, but it is not known whether rank $\Omega(N)$ is necessary
for solving mutual exclusion with $\O(1)$ RMRs per passage.
Examples of primitives that have rank $2N$ or more include an $r$-bounded fetch-and-increment
(i.e., $\phi(old, input) = \min(r-1, old + 1)$)
for $r \geq 2N$, which has rank $r$, and fetch-and-store, which has infinite rank.
Compare-and-swap as well as test-and-set can also be modeled as fetch-and-$\phi$ primitives,
but both have rank only two.  

Any mutual exclusion algorithm that uses only compare-and-swap and registers
requires $\Omega(\log N)$ RMRs \cite{atti:rmr,ghhw:cas}.
In contrast, there are mutual exclusion algorithms that use only fetch-and-store and
registers that require only $O(1)$ RMRs (e.g., \cite{craig:queue}).
So, from the point of view of supporting RMR-efficient implementations of mutual exclusion,
fetch-and-store is more powerful than compare-and-swap.
It is interesting that the opposite is the case from the point of view of supporting
wait-free implementations of objects.  It is well-known from Herlihy's work
that compare-and-swap and registers support wait-free implementation of any object
shared by any number of processes, while there are objects shared by only
three processes that cannot be implemented wait-free using only
fetch-and-store and registers \cite{herl:wait}.

\section{Road Map \label{sec_roadmap}}
First, we present the model of computation in Section~\ref{sec_mcd}.
In Section~\ref{sec_genform} we present our generic formulation of the
queue-based mutual exclusion algorithm, and prove its correctness
properties, assuming a suitable implementation of \MQ,
a novel queue-like data structure.
Then, in Sections \ref{sec_fi} and \ref{sec_fs},
we discuss two implementations of \MQ, based
on the fetch-and-increment and fetch-and-store primitives, respectively.
Our implementations closely follow the queuing code embedded in 
existing queue lock algorithms \cite{tand:spin,craig:queue}.
We conclude in Section~\ref{sec_conc} with a discussion of the
applicability of our analysis technique.


\newcommand{\States}{\ensuremath{\mathcal{S}}}
\newcommand{\sinit}{\ensuremath{s_{init}}}
\newcommand{\Ops}{\ensuremath{\mathcal{O}}}
\newcommand{\Resps}{\ensuremath{\mathcal{R}}}
\newcommand{\Procs}{\ensuremath{\mathcal{P}}}
\newcommand{\Vars}{\ensuremath{\mathcal{V}}}
\newcommand{\Base}{\ensuremath{\mathcal{B}}}
\newcommand{\Hists}{\ensuremath{\mathcal{H}}}
\newcommand{\Lin}{\ensuremath{\mathit{Lin}}}
\newcommand{\Last}{\ensuremath{\mathit{Last}}}
\newcommand{\Proc}{\ensuremath{\mathit{Proc}}}
\newcommand{\Queue}{\ensuremath{\mathit{Queue}}}
\newcommand{\myIndex}{\ensuremath{\mathit{myIdx}}}
\newcommand{\prevIndex}{\ensuremath{\mathit{prevIdx}}}
\newcommand{\tempIndex}{\ensuremath{\mathit{tempIdx}}}
\newcommand{\tempId}{\ensuremath{\mathit{tempId}}}
\newcommand{\invv}{\ensuremath{\beta}}
\newcommand{\invw}{\ensuremath{\gamma}}
\newcommand{\T}{\ensuremath{\tau}}
\newcommand{\B}{\ensuremath{\mathcal{B}}}
\renewcommand{\gets}{:=}

\newpage
\section{Model of Computation and Definitions \label{sec_mcd}}
Our model of computation is based on \cite{her:lin}.
A concurrent system models an asynchronous shared memory system
    where $N$ \emph{processes} communicate by executing \emph{operations}
    on shared \emph{objects}.
Formally, a concurrent system is represented as a triple $S = (\Procs, \Vars, \Hists)$,
    where $\Procs = \bc{0, 1, \ldots, N-1}$ is a set of process identifiers,
    $\Vars$ is a set of shared objects, also referred to as \emph{variables},
    and $\Hists$ is a set of \emph{execution histories}.
Each process identifier corresponds to a process, which is a sequential thread of control
	that invokes operations on objects, one at a time, and receives corresponding responses.
An object represents a data structure with a well-defined set of states
	and set of operations that modify the state and return responses to processes.
Processes and objects can be formally modelled as input/output automata \cite{lt:ioaut},
    but here we adopt a more informal approach.

\bigskip
\noindent \textbf{Steps} \ \\
Informally, we think of the behaviour of processes in a concurrent system $S = (\Procs, \Vars, \Hists)$
    as a collection of \emph{steps}. 
There are two categories of steps -- \emph{atomic} and \emph{non-atomic}.
In an atomic step, a process $p \in \Procs$ applies operation $op$ on some object
    $v \in \Vars$ and receives the response $ret$ of this operation.
This is denoted by a tuple $\br{\ATM, p, v, op, ret}$.
We use atomic steps to denote operations on atomic objects, such as those
   provided in hardware. 
In a non-atomic step, a process $p$ either invokes an operation $op$ on some object
    $v \in \Vars$, or it receives the response $ret$ of the last operation $p$
    invoked on $v$.
The former is called an \emph{invocation step}, and is represented by a tuple
    $\br{\INV, p, v, op}$.
The latter is called a \emph{response step}, and is represented by a tuple
    $\br{\RES, p, v, ret}$.
We use non-atomic steps (along with atomic steps)
    to denote operations on objects that are simulated
    in software from atomic objects, as explained later.

\bigskip
\noindent \textbf{Execution Histories} \ \\
    An execution history, or history for short, 
    is a sequence of steps.
An execution history is generated as processes accesses objects
    according to the transition functions of the corresponding
    automata, which we will describe using pseudocode.
The histories we will consider contain either only atomic steps,
    or a combination of atomic and non-atomic steps where
    each object is accessed by steps of exactly one category.

We say that $H$ is a history \emph{of} (or \emph{over}) object $v$ if every step in $H$ 
accesses $v$.
A response step $e_R = \br{\RES, p, v, \wild}$ in $H$ \emph{matches}
the last preceding invocation step $e_I = \br{\INV, p, v, \wild}$ in $H$ (if one exists).\footnote{
Here and in the remainder of the paper, ``$\wild$'' denotes a wildcard value.}
An invocation step is \emph{pending} in $H$ if it is not followed by a matching response step.

An \emph{operation execution} in a history $H \in \Hists$ is either a pair of matching
invocation/response steps, or a pending invocation step.  We call an operation execution
\emph{complete} in the former case, and \emph{pending} in the latter.
Two operation executions are \emph{concurrent} in $H$ unless the
response of one precedes the invocation of the other in $H$.
We say that $H$ is \emph{sequential} if it contains no concurrent operation executions,
and \emph{complete} if it contains no pending invocations.
The set $\Hists$ is \emph{prefix-closed}, meaning that if $H \in \Hists$ and $G$ is a prefix
of $H$ then $G \in \Hists$.

For every history $H$ and set $P$ of process IDs,
    we denote by $H|P$ the maximal subsequence
	of $H$ consisting only of steps by processes in $P$.
Similarly, for every history $H$ and set $V$ of objects,
    we denote by $H|V$ the maximal
    subsequence of $H$ consisting only of steps on objects in $V$.
For a single process ID $p$ or object $v$, we use $H|p$ and $H|v$ as shorthands for
$H|\bc{p}$ and $H|\bc{v}$, respectively.
A process $p$ is \emph{active} in a history $H$ if $H|p$ is not empty.

\bigskip
\noindent \textbf{Object Types and Conformity to a Type} \ \\
Every object has a \emph{type}    
    $\T = (\Procs, \States, \sinit, \Ops, \Resps, \delta)$ where $\Procs$ is a set of process IDs
	(defined as for concurrent systems), 
$\States$ is a set of states, $\sinit \in \States$ is the initial state, $\Ops$ is a set of operations,
$\Resps$ is the set of operation responses, and 
$\delta: \Procs \times \States \times \Ops \goesto \States \times \Resps$ is a (one-to-many) state transition mapping.
The transition mapping $\delta$ is intended to capture the behaviour of objects of type $\T$,
    in the absence of concurrency, as follows:
    if a process $p$ applies operation $op$ to an object
    of type $T$ that is in state $s$, then the object may return to $p$ the
    response $r$ and change its state to $s'$ if and only if $(s', r) \in \delta(p, s, op)$.
A complete, sequential execution history $H$ of object $v$ of type $\T$ induces a sequence of
    tuples $(p_i, op_i, r_i)$ such that in the $i$'th
    atomic step or operation execution in $H$ (depending on the structure of $H$),
	process $p_i$ applies operation $op_i$ and receives response $r_i$.
We say that $v$ \emph{conforms} to $\T$ in $H$ if there exists a sequence
    $s_0, s_1, s_2, \ldots$ of states of $\T$
    such that $s_0 = \sinit$ and for each $i \geq 1$,
    $(s_i, r_i) \in \delta(p_i, s_{i-1}, op_i)$.

\bigskip
\noindent \textbf{Algorithms} \ \\
An algorithm is a concurrent system $S = (\Procs, \Vars, \Hists)$
  where every history $H \in \Hists$ contains only atomic steps over $\Vars$.
We call such a history a \emph{one-level} history, to distinguish it from
  the more complex execution history of an \emph{implementation}, defined later.
The set of histories is defined informally through a pseudo-code procedure for
   each process as follows:
For each operation that a process $p$ applies to a shared variable,
    $H$ records an atomic step that encodes the variable, the operation applied,
    and its response.
Accesses to private variables correspond to state changes in the automaton
    for a process and are not explicitly recorded in the history.
Steps of different processes can be interleaved in $H$ arbitrarily.
An infinite history $H$ of an algorithm is \emph{fair}
  if every process that is active in $H$ takes infinitely many steps.
(We do not consider terminating algorithms in this paper.)

\bigskip
\noindent \textbf{Implementations} \ \\
An \emph{implementation} describes how to simulate a \emph{target object}
    of a particular \emph{target type}
    using a set of \emph{base objects} of specified types.
Specifically, for each operation of the target type and each process, 
    we define an \emph{access procedure} that computes the response of
    the operation under consideration by performing operations on the base objects.
An implementation is a concurrent system denoted
    $I = (\Procs, \Vars, \Hists)$ where the set of shared objects $\Vars$
    consists of a distinguished target object, denoted $T$, and a set of base objects.
Histories in $\Hists$ contain a combination of atomic and non-atomic steps.
Every history $H \in \Hists$ is \emph{well-formed}, meaning that   
 the following conditions hold:
\begin{itemize}
\item $T$ is accessed only using non-atomic steps, and for every base object
      $v \in \Vars$, $v$ is accessed only using atomic steps.
\item For every base object $v \in \Vars$, $v$ conforms to its type in $H|v$.
\item If $e_I = \br{\INV, p, T, \wild}$ is pending in $H$ then $e_I$ is
	the last non-atomic step performed by $p$ in $H$.
\item If $e_R = \br{\RES, p, T, \wild}$ is in $H$ then $e_R$ matches
      the last invocation step of $p$ that precedes $e_R$ in $H$.
\item If $e_A = \br{\ATM, p, \wild, \wild, \wild}$ is in $H$ then
        it occurs after some invocation step and before the matching response
        (if one exists).
\end{itemize}
We call an execution history of an implementation a \emph{two-level} history
    since operation executions on the target object and on base objects are nested.

Histories in $\Hists$ correspond to executions of the access procedures as follows.
When a process $p$ begins executing the access procedure for operation $op$ on $T$,
	the history records the step $\br{\INV, p, T, op}$.
As $p$ subsequently executes the access procedure, the history
	records corresponding atomic steps by $p$ on base objects.
Finally, when the access procedure returns a value $ret$, then the history
	records the response step $\br{\RES, p, T, ret}$.
Processes may call the access procedures arbitrarily many times and in arbitrary order.
An infinite history $H$ of an implementation is \emph{fair} if every process that is
  active in $H$ either takes infinitely many steps, or applies
  a response step as its last step in $H$.
Informally, this means that in a fair history a process may not stop executing
  in the middle of an access procedure.
  
\bigskip
\noindent \textbf{Linearizability} \ \\
Linearizability \cite{her:lin} is widely accepted as a correctness condition for
concurrent objects.  Informally, it states that operation executions
in a history of an implementation must appear to take effect instantaneously at some point
between the corresponding invocation and response steps.
Formally, linearizability is defined as follows.
Given a history $H$ of an implementation, $<_H$ is the partial order over the set of
operation executions in $H$ defined as follows: $oe_1 <_H oe_2$ iff the response of 
$oe_1$ occurs in $H$ before the invocation of $oe_2$.
Two execution histories $G$ and $H$ are \emph{equivalent} if
    every process executes the same sequence of steps in both histories.
Letting $T$ denote the target object,
a \emph{completion} of $H|T$ is a well-formed history $H'$ obtained from $H$
by either completing (with a response event) or removing every pending operation execution.
$H|T$ is \emph{linearizable with respect to type $\T$} if it has a completion equivalent
  to some complete sequential history $\L{H}$ over $T$ such that $<_H \subseteq <_{\L{H}}$
and where $T$ conforms to type $\T$ in $\L{H}$.
In this case we say that $\L{H}$ is a \emph{linearization} of $H$.
We denote the set of possible linearizations of $H$ by $\Lin(H)$.
We say that an implementation $I = (\Procs, \Vars, \Hists)$ is \emph{linearizable
  with respect to type $\T$} if for every history $H \in \Hists$,
  $H|T$ is linearizable with respect to type $\T$.

\bigskip
\noindent \textbf{Additional Notation} \ \\
Let $G, H$ be execution histories.
If $s$ is a step, we denote by $s \in H$ that step $s$ occurs in $H$,
by $proc(s)$ the process that executes $s$, and by $var(s)$ the object on which $s$ operates.
We denote by $G \preceq H$ that $G$ is a prefix of $H$, and by $G \prec H$ that $G$ is a proper prefix of $H$.
If $v$ is an object and $H$ is an execution history such that $H|v$ is complete and sequential, then we denote the state of $v$ at the end of $H|v$ by $v^H$.

Given execution histories (or, more generally, sequences) $H$ and $G$,
let $G \circ H$ denote the concatenation of $G$ and $H$ (i.e., elements of $H$ appended to $G$).
If $G$ is finite, $|G|$ denotes the length of $G$.
For $0 \leq i < |G|$, $G[i]$ denotes the $i$'th step (counting from 0) of $G$.
$G[i .. j]$ denotes the subsequence of $G$ consisting of all $G[k]$ such that $i \leq k \leq j$.


\newpage
\section{Generic Queue-Based Algorithm \label{sec_genform}}
\subsection{The \MQ\ Type \label{sec_sup_ds}}
An $N$-process \MQ\ is a queue-like object type that stores a subset of $N$ process IDs
  (subsequently also referred to as processes).
The state of \MQ\ is an ordered pair $(Q, V)$, where $Q$ and $V$ are
a sequence and a set, respectively, of elements from $\Procs$.
Informally, $Q$ represents the sequence of processes waiting to enter the critical
section, and $V$ is a subset of these processes that are \emph{visible}.
Intuitively, a process becomes visible when it ``makes itself known'' to its predecessor
in the queue. 
The initial state is $(\bg{}, \emptyset)$.
In addition, we define a special \emph{broken} state $\bot$,
indicating that a process has violated the etiquette for accessing 
\MQ\ (explained below).

A \MQ\ supports three types of operations: $\enqueue$, $\isHead$, and $\dequeue$.
Informally, $\enqueue$ adds the executing process to the end of the queue and
always returns the response $\OK$;
$\isHead$ makes the executing process visible and returns $\true$ if and only if
this process is the head of the queue; 
and $\dequeue$ removes the executing process from the head of the queue, and returns
the ID of the successor process in the queue, if it exists and is visible (or $-1$ otherwise).
As mentioned earlier, processes are expected to follow a certain etiquette in accessing
\MQ.  Specifically, a process must not invoke $\enqueue$ if it is already in the queue,
$\isHead$ if it is not in the queue or is already visible, and
$\dequeue$ if it is not the head of the queue or is not visible.
Failure to comply with this etiquette causes the \MQ\ to enter the broken state $\bot$, and
thereafter all responses are completely arbitrary.
These restrictions on accessing \MQ\ make it easier to implement this object.
As we will see, they are observed by our generic algorithm that uses \MQ\ to solve mutual exclusion
(see Section~\ref{sec_mutexalgo}).


Prima facie, it would seem that we can have a simpler definition
of \MQ, and a correspondingly simpler version of the generic mutual exclusion algorithm
based on \MQ, by combining the $\enqueue$ and $\isHead$
operations into a single operation that adds the ID of the executing process
to the end of the queue, and returns \true\ if that process is the head of the
queue and \false\ otherwise.  Unfortunately, the resulting operation
seems too strong;  we were not able to find an implementation for
it that uses standard synchronization primitives and incurs only
a constant number of RMRs.  By splitting the functionality into
two separate operations, such implementations become feasible.

Formally, an $N$-process \MQ\ is specified by the tuple $(\Procs, \Ops, \Resps, \States, \tau)$ where 
\begin{eqnarray*}
 \Procs &=& \bc{0, 1, \ldots, N-1} \\
 \Ops &=& \bc{\enqueue,\,\isHead,\,\dequeue} \\
 \Resps &=& \bc{\true, \false, -1} \cup \bc{0, 1, \ldots, N-1} \\
 \States &=& \bc{\bot} \cup \left\{(Q, V) \mid Q \mbox{ is a permutation of a subsequence of } \bg{0, 1, \ldots, N-1}\right.\\
   && \tab \tab \ \ \ \ \ \ \left.\mbox{ and } p \in V \mbox{ only if } p \in Q\right\}
\end{eqnarray*}
and the state transition mapping $\tau$ is defined as follows:
\begin{tabbing}
    $\tau(p, s, \enqueue)$ \\
    \tab \=
    $= \twodefo{\bc{((Q \circ \bg{p}, V), \OK)}}{s = (Q, V) \mbox{ and } p \not\in Q}
         {\bc{(\bot, \rt) \mid \rt \in \Resps}}$ \\
\end{tabbing}

\begin{tabbing}
    $\tau(p, s, \isHead)$ \\
    \tab \= $= \threedefo
      {\bc{((Q, V \cup \bc{p}), \true)}}{s = (Q, V), p \in Q, p\not\in V \mbox{ and } Q[0] = p}
      {\bc{((Q, V \cup \bc{p}), \false)}}{s = (Q, V), p \in Q, p\not\in V \mbox{ and } Q[0] \neq p}
      {\bc{(\bot, \rt) \mid \rt \in \Resps}}$\\
\end{tabbing}

\begin{tabbing}
    $\tau(p, s, \dequeue)$ \\
    \tab \= $= \threedefo
       {\bc{((Q[1..|Q|-1], V \setminus \bc{p}), Q[1])}}{s = (Q, V), p \in Q, p \in V, Q[0] = p, \\ & \ \ \ \ |Q| > 1 \mbox{ and } Q[1] \in V}
       {\bc{((Q[1..|Q|-1], V \setminus \bc{p}), -1)}}{s = (Q, V), p \in Q, p \in V, Q[0] = p, \\ & \ \ \ \ (|Q| = 1 \mbox{ or } Q[1] \not\in V)}
       {\bc{(\bot, \rt) \mid \rt \in \Resps}}$
\end{tabbing}

%
%

\bigskip
\begin{observation}  \label{o_dupno}
Let $H$ be an execution history over an atomic $N$-process \MQ\ object $M$ such that
$M^H = (Q, V) \neq \bot$. Then the following hold:
\begin{enumerate}\renewcommand{\labelenumi}{(\alph{enumi})}
\item for every process $p$, if $p \in V$ then $p \in Q$ 
\item for every process $p$, $Q$ contains at most one instance of $p$
\end{enumerate}
\end{observation}

\bigskip
Given a state $s = (Q, V)$ of a \MQ\ object, $s \neq \bot$,
we define the following predicates and functions.
    \begin{eqnarray*}
        QProcs(s) &:=& \bc{p \mid p \in Q}\\
        V\!isProcs(s) &:=& V \\
        empty(s) &:=& \twodefo{\true}{Q = \bg{}}{\false} \\
        head(s)    &:=& \twodefo{Q[0]}{|Q| \geq 1}{\perp}  \\
        pred(s, p) &=& \otwodef{q}{\bg{q, p} \mbox{ is a subsequence of } Q}{\bot}{\mbox{$Q$ has no such subsequence}} \\
        succ(s, p) &=& \otwodef{q}{\bg{p, q} \mbox{ is a subsequence of } Q}{\bot}{\mbox{$Q$ has no such subsequence}}
    \end{eqnarray*}
Note that for every $p \in \Procs$, the values $pred(s, p)$ and $succ(s, p)$
are uniquely defined by Observation~\ref{o_dupno} (b).
If $p,q \in \Procs$ and $s$ is a \MQ\ state then we use the phrases
``$s$ is empty,'' ``$p$ is in the queue,'' ``$p$ is the head of $s$,'' ``$p$ is visible in $s$,''
``$p$ is the successor of $q$ in $s$'' and ``$p$ is the predecessor of $q$ in $s$,''
to denote the conditions $empty(s)$, $p \in QProcs(s)$, $p = head(s)$, $p \in V\!isProcs(s)$,
$p = succ(s, q)$, and $p = pred(s, q)$, respectively.

\subsection{Generic Mutual Exclusion Algorithm \label{sec_mutexalgo}}
In this section we analyze the mutual exclusion algorithm shown in
Figure~\ref{fig_genalg}.  In addition to an atomic \MQ\ object,
    the algorithm uses an array $\Wayte[00..N-1]$ of Boolean read/write registers.

Informally, the algorithm uses the \MQ\ object $M$ to maintain a queue of
processes that are competing to enter the critical section.
The $\enqueue$ operation at line~\ref{gm_2} constitutes the doorway, and the remaining statements
leading up to the CS comprise the waiting room.  In the exit protocol, spanning
lines 7 to 9, a process signals its successor in $M$ (if present and visible) to exit
the waiting room and proceed to the CS.

We use syntax of the form $V.$\texttt{op}$(args)$ in Figure~\ref{fig_genalg} to indicate
that process $p$ invokes operation \texttt{op}$(args)$ on the shared variable $V$.
Operations on shared registers are denoted $\opread$ and $\opwrite$.

\begin{figure}
  \begin{function}[H]
	\begin{tabbing}
Shared variables:\\
 	\tab \= $\Wayte$: \hspace{0.7cm} \= array $[0..N-1]$ of Boolean, initially all $\true$ \\
	\> \> ($\Wayte[p]$ local to $p$ on a DSM machine) \\
	\> $M$: \> $N$-process \MQ\  \\
Private per-process variables:\\
	\> $nextHead$: \> integer $-1 .. N-1$  \;
	\end{tabbing}
Algorithm for process $p$:   \; \Indp\Indp 
  \SetKwBlock{LoopForever}{loop}{forever}
\LoopForever {
\lnl{gm_1}     NCS  \;
\lnl{gm_2}     $M.\enqueue$ \;
\lnl{gm_3}    \If{$\neg M.\isHead$}{
\lnl{gm_4}    \While{$\Wayte[p].\opread = \true$}{} 
\lnl{gm_5}     $\Wayte[p].\opwrite(\true)$  \;
      	       }
\lnl{gm_6}    CS \\
\lnl{gm_7}     $nextHead \gets M.\dequeue$ \;
\lnl{gm_8}     \If{$nextHead \neq -1$}{
\lnl{gm_9}    $\Wayte[nextHead].\opwrite(\false)$ \; 
      	      }
}
\end{function}
\caption{Algorithm~GQME (Generic Queue-based Mutual Exclusion) for $N$ processes.}
\label{fig_genalg}
\end{figure}

\subsubsection{Correctness Properties}
\noindent
Mutual Exclusion (ME): at most one process is in the CS at any time.

\medskip
\noindent
First-Come First-Served (FCFS): processes enter the CS in the order in which they
are enqueued at line~\ref{gm_2}.

\medskip
\noindent
Lockout Freedom (LF): if a process leaves the NCS then it eventually enters the CS.

\medskip
\noindent
Bounded Exit (BE): if a process leaves the CS then it enters the NCS within a bounded number
of its own steps. 

\subsubsection{Proof of Correctness}\label{sec_poc_gqme}
Let $S = (\Procs, \Vars, \Hists)$ be the concurrent system  
    corresponding to Algorithm~GQME where
    $\Procs = \bc{0, 1, \ldots, N-1}$, $\Vars = \bc{M, \Wayte[0], \ldots \Wayte[N-1]}$,
    and $\Hists$ is the set of execution histories of Algorithm~GQME.
Each (concurrent) execution of Algorithm~GQME is represented by
    a one-level history $H \in \Hists$ as follows.
For each operation that a process $p$ applies to a shared variable,
    (e.g., $M.\enqueue$ at line~\ref{gm_2}),
    $H$ records an atomic step.\footnote{
Note that $H$ does not record steps corresponding to the private variable $nextHead$.
The value of $nextHead$ is part of the local state of a process.}
The sequence of steps of each process in $H$
	is determined by the pseudocode shown in Figure~\ref{fig_genalg}.
For example, if process $p$ applies operation $\isHead$ to $M$
    (see line~\ref{gm_3}) with response $\false$,
	then the next step of $p$ in $H$ (if one exists)
	applies $\opread$ to $\Wayte[p]$;
	otherwise, the next step of $p$ in $H$ (if one exists),
	applies $\dequeue$ to $M$.
The steps of different processes can be interleaved in $H$ in any way
    provided that each variable in $\Vars$ conforms to its type in $H$.

For any history $H \in \Hists$, any process $p \in \Procs$, and any integer $i \in \Z^+$,
we say that $p$ is \emph{in the CS in passage $i$} at the end of $H$
    if and only if $p$ performs its last step in $H$ during its $i$'th passage through Algorithm~GQME,
    and furthermore this step is:
    either $\bg{(\ATM, p, M, \isHead, \true)}$ (see line~\ref{gm_3});
    or $\bg{(\ATM, p, \Wayte[p], \opwrite(\true), \OK)}$ (see line~\ref{gm_5}).
Similarly, we say that $p$ has \emph{completed the CS in passage $i$} at the end of $H$
    if and only if $H$ contains a step $(\ATM, p, M, \dequeue, -)$
    (see line~\ref{gm_7}) performed by $p$ during passage $i$ through Algorithm~GQME.

\begin{figure}
    \vspace{1pc}\hspace{3pc}
    \entrymodifiers={++[F-]}
    \turnradius={1.5pc}
    \xymatrix@+1.7pc{
        *\txt{} & \txt{NEAR\_NCS} \ar@(u,u) \ar[d] |*+{M.\enqueue} & *\txt{} & *\txt{} \\
        *\txt{} & \txt{DOORWAY} \ar[dl] |*+\txt{$M.\isHead$ \\ returns $\true$}
                                \ar[dr] |*+\txt{$M.\isHead$ \\ returns $\false$}  & *\txt{} & *\txt{} \\
        \txt{NO\_WAIT} \ar `d[dr] [dr] & *\txt{} &  \txt{WAIT} \ar[d] |*+{\Wayte[p] \longleftarrow \true} &  *\txt{} \\
        *\txt{} & *+{\mbox{ }} \ar @{>->} `d[dr] `[urr] |*++{M.\dequeue}  `[uuu] [uuu]   & \txt{DONE\_WAIT} \ar[l] & *\txt{} \\
        *\txt{} & *\txt{ }  & *\txt{} & *\txt{}
    }
    \caption{Phase transitions of process $p$ executing Algorithm~GQME.}
    \label{fig_ptd}
    \end{figure}
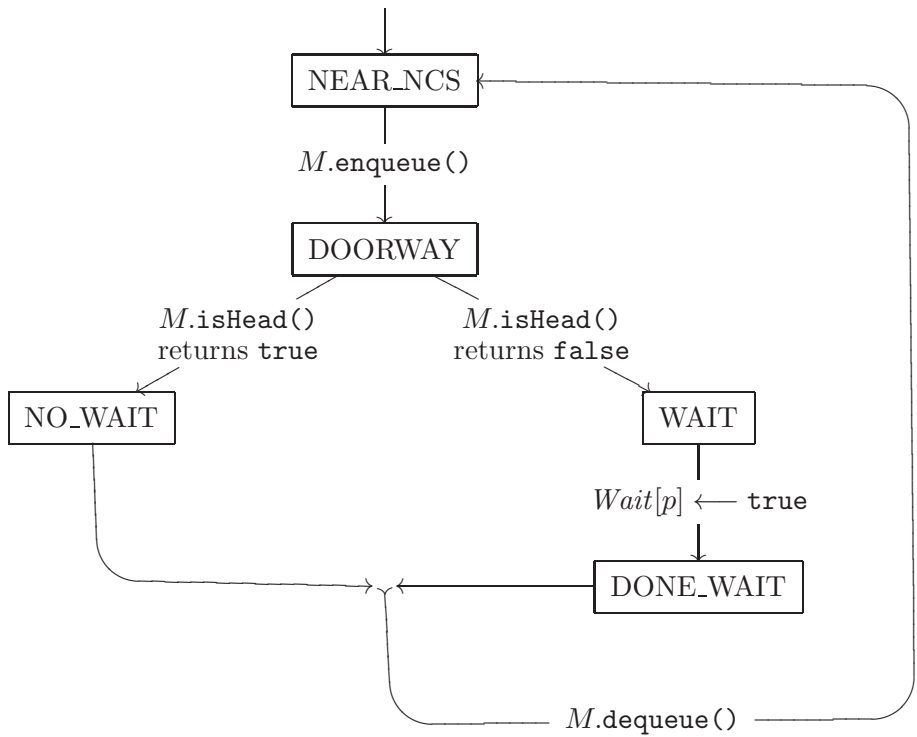

For ease of exposition, we distinguish a number of phases in which a process may be 
at the end of a history $H \in \Hists$.
The phases are defined in Table~\ref{tab_ppd} and the transitions between them
are illustrated in Figure~\ref{fig_ptd}.
Each phase is bounded by steps on shared objects.

Note that the first five phases defined in Table~\ref{tab_ppd} are mutually exclusive,
whereas EXIT is a sub-phase of NEAR\_NCS, and is not necessarily traversed by a process in every passage
    through Algorithm~GQME.
We will subsequently use the name of a phase as a predicate indicating that a process
is in the given phase, e.g., WAIT$(p)^H = \true$ iff process $p$ is in the WAIT phase
at the end of a history $H \in \Hists$.

    \begin{table}
    \begin{center}
    \begin{tabular}{|c|c|c|c|}
    \hline
    Phase name  & From operation    & To operation      & Notes/conditions  \\\hline\hline
    DOORWAY     & \enqueue\ at line~\ref{gm_2}  & \isHead\ at line~\ref{gm_3} & \\
    WAIT        & \isHead\ at line~\ref{gm_3}   & \opwrite\ at line~\ref{gm_5} & no branch at line~\ref{gm_3} \\
    DONE\_WAIT  & \opwrite\ at line~\ref{gm_5} & \dequeue\ at line~\ref{gm_7}    &  \\
    NO\_WAIT    & \isHead\ at line~\ref{gm_3} & \dequeue\ at line~\ref{gm_7} & branch from line~\ref{gm_3} to line~\ref{gm_6}  \\
    NEAR\_NCS   & \dequeue\ at line~\ref{gm_7}  & \enqueue\ at line~\ref{gm_2}  & via the NCS \\ \hline
    EXIT        & \dequeue\ at line~\ref{gm_7}  & \opwrite\ at line~\ref{gm_9} & no branch at line~\ref{gm_8} \\
    \hline
    \end{tabular}
    \caption{Process phase definitions.}
    \label{tab_ppd}
    \end{center} \end{table}

\newpage
To prove the correctness of the algorithm we will establish the following invariant.

\begin{invariant}\label{a0_i1}
Let $H \in \Hists$.
Define $lastPred(H, p)$ as the last process enqueued before $p$'s last
$\enqueue$ operation in $H$, or $\bot$ if no such process exists.
Then $M^H \neq \bot$ and for every $p \in \Procs$,
the following statements hold, collectively denoted Invariant~\ref{a0_i1}--$(H, p)$:
\end{invariant}

    \begin{enumerate}\renewcommand{\labelenumi}{(\alph{enumi})}
    \item  \hspace{0.5cm} if \ $p = head(M^{H})$ \ then
    \begin{tabbing}
    \tab \=
    NEAR\_NCS$(p)^H$ \hspace{1pc}  \= \ \ \ $=$ \ \ \ \= $\false$   \\
    \> DOORWAY$(p)^H$  \> $\imps$ \> $\Wayte[p]^{H} = \true$ \\
    \> WAIT$(p)^H$  \> $\imps$ \> $\Wayte[p]^{H} = \twodefo{\true}{lastPred(H, p) \neq \bot \we \\ & \mbox{EXIT}(lastPred(H, p))^H}{\false}$ \\
    \> DONE\_WAIT$(p)^H$ \> $\imps$ \> $\Wayte[p]^{H} = \true$ \\
    \>  NO\_WAIT$(p)^H$ \> $\imps$ \> $\Wayte[p]^{H} = \true$
    \end{tabbing}
    \item \hspace{0.5cm} if \ $p \in QProcs(M^{H}) \wedge p \neq head(M^{H})$ \ then
    \begin{tabbing}
    \tab \=  DOORWAY$(p)^H$  \ \   \=$=$  \ \ \=  $\twodefo{\true}{p \not\in V\!isProcs(M^{H})}{\false}$ \\
    \> WAIT$(p)^H$  \> $=$ \> $\twodefo{\true}{p \in V\!isProcs(M^{H})}{\false}$ \\
    \> $\Wayte[p]^{H}$ \> $=$ \> $\true$
    \end{tabbing}
    \item \hspace{0.5cm}  if \ $p \not\in QProcs(M^{H})$ \ then
    \begin{tabbing}
    \tab \= NEAR\_NCS$(p)^H$ \= $=$ \= $\true$ \\
    \> $\Wayte[p]^{H}$ \> $=$ \> $\true$
    \end{tabbing}
    \end{enumerate}
\vspace{1pc}
\renewcommand{\labelenumi}{\arabic{enumi}.}

\begin{theorem} \label{i0_t1}
For any $H \in \Hists$, Invariant~\ref{a0_i1} holds for $H$.
\begin{proof}
We proceed by induction on $|H|$.

\medskip
\noindent\textbf{Basis:} $|H| = 0$.
In this case, $M^H$ is the initial state $(\bg{}, \emptyset)$, hence $M^H \neq \bot$.
Since $empty(M^H)$ is true, for every $p \in \Procs$
parts (a) and (b) of Invariant~\ref{a0_i1}--$(H, p)$ hold trivially (since their antecedents are false),
and part (c) holds because NEAR\_NCS$(p)^H$ and $\Wayte[p]^{H} = \true$ by initialization.  

\medskip
\noindent\textbf{Induction Hypothesis:}
For any $k > 0$, assume Theorem~\ref{i0_t1} holds for all 
histories $H \in \Hists$ such that $|H| < k$.

\medskip
\noindent\textbf{Induction Step:} 
We must prove Theorem~\ref{i0_t1} for all $H$ such that $|H| = k$.
Let $\sigma$ be the last step in $H$ and let
$G$ satisfy $H = G \circ \sigma$.
By the IH, $M^G \neq \bot$ and Invariant~\ref{a0_i1}--$(G, p)$ holds for all $p$.
Define a \emph{critical} operation as a write operation to an element of $\Wayte$
or any operation on $M$ (i.e., an operation causing a process to change phases).
If $\sigma$ is not critical, the fact that Theorem~\ref{i0_t1} holds for $G$
immediately implies that it also holds for $H$.
Consequently, it suffices to prove that Theorem~\ref{i0_t1} holds for $H$ if 
$\sigma$ is a critical step. 
We proceed by cases on $\sigma$.

\medskip
\noindent\textbf{Case A:} step $\sigma$ is an $M.\enqueue$ by $p$
(see line~\ref{gm_2}).  In this case, $p$ goes from NEAR\_NCS to DOORWAY.
Since $M^G \neq \bot$ by the IH and NEAR\_NCS$(p)^G$,
Invariant~\ref{a0_i1}--$(G, p)$ implies $p \not\in QProcs(M^G)$,
and $M^H \neq \bot$ holds by the state transition relation of \MQ.
Next, note that for every $q \in \Procs \setminus \bc{p}$, Invariant~\ref{a0_i1}--$(G, q)$
implies Invariant~\ref{a0_i1}--$(H, q)$.
It remains to show Invariant~\ref{a0_i1}--$(H, p)$. \\
\textbf{Subcase A1:} $p = head(M^{H})$.
Since NEAR\_NCS$(p)^G$, Invariant~\ref{a0_i1}--$({G}, p)$ implies that
$\Wayte[p]^{G} = \true$.  Thus, $\Wayte[p]^{H} = \true$, and part (a) of
Invariant~\ref{a0_i1}--$({H}, p)$ holds.  
Parts (b) and (c) follow trivially since $p = head(M^{H})$. \\
\textbf{Subcase A2:} $p \neq head(M^{H})$.
We have $\Wayte[p]^{G} = \true$ as in subcase A1.
Since $p \not\in V\!isProcs(M^{H})$ by $\sigma$,
part (b) of Invariant~\ref{a0_i1}--$({H}, p)$ holds.  
Parts (a) and (c) follow trivially since $p \in QProcs(M^{H})$ and $p \neq head(M^{H})$.

\medskip
\noindent\textbf{Case B:} step $\sigma$ is an $M.\isHead$ by $p$,
with response $\rt$ (see line~\ref{gm_3}).  In this case, $p$ goes from DOORWAY to WAIT or NO\_WAIT. 
Since DOORWAY$(p)^G$, Invariant~\ref{a0_i1}--$(G, p)$ implies $p \in QProcs(M^{G})$.
Moreover, since $M^G \neq \bot$ by the IH and $G|M|p$ ends with an $\enqueue$ step,
it follows that $p \not\in V\!isProcs(M^{G})$ and $M^H \neq \bot$.
Furthermore, $\rt$ is either $\true$ or $\false$ by the specification of \MQ.
As in Case A, Invariant~\ref{a0_i1}--$({H}, q)$ holds for every $q \in \Procs \setminus \bc{p}$
and it remains to show Invariant~\ref{a0_i1}--$({H}, p)$. \\
\textbf{Subcase B1:} the last step in $H$ (an $M.\isHead$) returns $\true$.
Then $p = head(M^{H})$ by the specification of \MQ\ (since $M^G \neq \bot$)
and NO\_WAIT$(p)^H$ is true by the algorithm. 
From Invariant~\ref{a0_i1}--$({G}, p)$ part (a) we have that $\Wayte[p]^{G} = \true$, hence
$\Wayte[p]^{H} = \true$.
Thus, part (a) of Invariant~\ref{a0_i1}--$({H}, p)$ holds.
Parts (b) and (c) hold trivially since $p = head(M^{H})$. \\
\textbf{Subcase B2:} the last step in $H$ (an $M.\isHead$) returns $\false$.
Then $p \neq head(M^{H})$ and $p \in V\!isProcs(M^{H})$
by the specification of \MQ\ (since $M^G \neq \bot$) and WAIT$(p)^H$ is true by the algorithm.
We have $\Wayte[p]^{H} = \true$ as in subcase B1; since $p \in V\!isProcs(M^{H})$,
part (b) of Invariant~\ref{a0_i1}--$({H}, p)$ holds.
Parts (a) and (c) hold trivially since  $p \neq head(M^{H})$ and $p \in V\!isProcs(M^{H})$. 

\medskip
\noindent\textbf{Case C:} step $\sigma$ is a write of $\true$ by $p$ to $\Wayte[p]$
(see line~\ref{gm_5}).  In this case, $p$ goes from WAIT to DONE\_WAIT.
It follows that $M^H \neq \bot$ since $M^H = M^G$ and $M^G \neq \bot$ by the IH.
As in the previous case, for all $q \in \Procs \setminus \bc{p}$,
Invariant~\ref{a0_i1}--$(G, q)$ immediately implies that Invariant~\ref{a0_i1}--$(H, q)$ holds,
and so it remains to show that Invariant~\ref{a0_i1}--$({H}, p)$ holds. 
Since WAIT$(p)^G$, part (c) of Invariant~\ref{a0_i1}--$({G}, p)$ implies that $p \in QProcs(M^{G})$.
Moreover, $\Wayte[p]^{G} = \false$ by the algorithm since $p$'s last read of $\Wayte[p]$ in
${G}$ returns $\false$ and the value of $\Wayte[p]$ is not changed until $\sigma$ occurs.
Since $\Wayte[p]^{G} = \false$, Invariant~\ref{a0_i1}--$({G}, p)$ implies $p = head(M^{G})$.
Thus, $p = head(M^{H})$ holds; furthermore, $\Wayte[p]^{H} = \true$ by the effect of 
the step $\sigma$.
This implies part (a) of Invariant~\ref{a0_i1}--$(H, p)$.
Parts (b) and (c) hold trivially since $p = head(M^{H})$.

\medskip
\noindent\textbf{Case D:} step $\sigma$ is an $M.\dequeue$ by $p$,
with response $\rt$
(see line~\ref{gm_7}).  In this case, $p$ goes from NO\_WAIT or DONE\_WAIT to NEAR\_NCS. 
Since either NO\_WAIT$(p)^G$ or DONE\_WAIT$(p)^G$,  Invariant~\ref{a0_i1}--$({G}, p)$ implies
that $p = head(M^{G})$.  
Moreover, since $G|M|p$ ends with an $\isHead$ step, and since 
$M^G \neq \bot$ by the IH, $\rt$ is either $-1$ or $succ(M^G, p)$ by the specification of \MQ,
and $p \in V\!isProcs(M^{G})$.  Thus, $M^H \neq \bot$ is true.
Now, let $s = succ(M^G, p)$, and note that either $s = \bot$ and $empty(M^H)$, or $s = head(M^H)$.
It follows that for every $q \in \Procs \setminus \bc{p, s}$, Invariant~\ref{a0_i1}--$({G}, q)$ 
implies Invariant~\ref{a0_i1}--$({H}, q)$. 
Next, consider Invariant~\ref{a0_i1}--$({H}, p)$.
To that end, we have $p \not\in QProcs(M^{H})$ by the specification of \MQ,
and $\Wayte[p]^{G} = \true$ by part (a) of Invariant~\ref{a0_i1}--$({G}, p)$
(since $p = head(M^{G})$, as argued above). Consequently,
part (c) of Invariant~\ref{a0_i1}--$({H}, p)$ holds, and parts (a) and (b) follow trivially. 
Finally, we must show Invariant~\ref{a0_i1}--$({H}, s)$ supposing that $s \neq \bot$.
Observe that $s \neq \bot$ implies $\neg empty(M^H)$ and $s = head(M^H)$ by $\sigma$.
Moreover, $p \neq s$ by Observation~\ref{o_dupno} (b) applied to $G$, so $s \neq head(M^{G})$.
\\
\textbf{Subcase D1:} step $\sigma$ (an $M.\dequeue$) returns $-1$.
It follows that $s \not\in V\!isProcs(M^H)$.
Consequently, by part(b) of Invariant~\ref{a0_i1}--$({G}, s)$ we have DOORWAY$(s)^G$
and $\Wayte[s]^{G} = \true$.  Thus, DOORWAY$(s)^H$ and $\Wayte[s]^{H} = \true$ hold, which implies
part(a) of Invariant~\ref{a0_i1}--$({H}, s)$.  In addition, parts (b) and (c) hold trivially. 
\\
\textbf{Subcase D2:} step $\sigma$ (an $M.\dequeue$) returns a process ID $\rt$.
It follows that $\rt = s$ and $s \in V\!isProcs(M^{H})$.
Consequently, by part (b) of Invariant~\ref{a0_i1}--$({G}, s)$ we have
WAIT$(s)^G$ and $\Wayte[s]^{G} = \true$.  Thus, WAIT$(s)^H$ and $\Wayte[s]^{H} = \true$.
Observe that EXIT$(p)^H$ holds by the algorithm, so part (a) of Invariant~\ref{a0_i1}--$({H}, s)$ holds.
In addition, parts (b) and (c) hold trivially. 

\medskip
\noindent\textbf{Case E:} step $\sigma$ is a write of $\false$ by $p$ to $\Wayte[i]$ 
for some $i$
(see line~\ref{gm_9}).  In this case, $p$ leaves EXIT (which is part of NEAR\_NCS) and remains in NEAR\_NCS.
As in Case C, it follows that $M^H \neq \bot$.
Let $D$ and $E$ be prefixes of $G$ such that
$E = D \circ \bg{\br{\ATM, p, M, \dequeue, i}}$ and $|D|$ is maximal.
Since $M^G \neq \bot$ it follows that $M^D \neq \bot$, $M^E \neq \bot$,
$p \in QProcs(M^{D})$, and $p = head(M^D)$.
Let $s = succ(M^D, p)$, and observe that, as in Case D, $p \neq s$, hence $s \neq head(M^{D})$.
Also note that $i \neq -1$ since $p$ has branched to line~\ref{gm_9}, hence $i = s$, $s \in V\!isProcs(M^D)$,
$s \in V\!isProcs(M^E)$, and $s = head(M^{E})$.  Since, $s \neq head(M^{D})$ and $s \in V\!isProcs(M^D)$,
part (b) of Invariant~\ref{a0_i1}--$(D, s)$ implies WAIT$(s)^D$ and $\Wayte[s]^D = \true$.
Next, note that $p = lastPred(E, s)$ and that $p$ performs no critical steps in $G$ after $E$.
Moreover, for every history $F$ such that $E \preceq F \preceq G$,
EXIT$(p)^F$ holds and so a straightforward induction on $|F|$ shows
(using part (a) of Invariant~\ref{a0_i1}--$(F, s)$)
that $s = head(M^F)$, $\Wayte[s]^{F} = \true$, WAIT$(s)^F$, and $p = lastPred(F, s)$.
Thus, $s = head(M^H)$, WAIT$(s)^H$, and $p = lastPred(H, s)$ all hold, 
and $\Wayte[s]^{H} = \false$ by the effect of step $\sigma$.
Since $\neg$EXIT$(p)^H$, part (a) of Invariant~\ref{a0_i1}--$({H}, s)$ is satisfied.
In addition, parts (b) and (c) hold trivially.
Finally, for every $q \in \Procs \setminus \bc{s}$, note that Invariant~\ref{a0_i1}--$(H, q)$
follows immediately from Invariant~\ref{a0_i1}--$(G, q)$.
\end{proof}
\end{theorem}

%
%
%

\begin{lemma} \label{i0_lem_fcfs}
Let $H \in \Hists$ and suppose that in $H$
  process $p$ executes $M.\enqueue$ in passage $i$ before $q$
executes $M.\enqueue$ in passage $j$, and at the end of which $q$
is in the CS in passage $j$.
Then $p$ has executed $M.\dequeue$ in passage $i$ in $H$.
\begin{proof}
Let $H$, $p$ and $q$ be as in the hypothesis of the lemma and suppose for contradiction
that $p$ has not executed $M.\dequeue$ in passage $i$ in $H$.
From Theorem~\ref{i0_t1} and Invariant~\ref{a0_i1}--$(H, q)$, it follows that 
$M^H \neq \bot$ and $q = head(M^H)$.
Since $p$ was enqueued in passage $i$ before $q$ in passage $j$, this implies that
$p$ in passage $i$ has been dequeued in $H$.
Since $M^H \neq \bot$, it follows that $p$ has executed $M.\dequeue$ in passage $i$ in $H$,
which contradicts the original hypothesis.
\end{proof}
\end{lemma}

\begin{corollary} \label{i0_me}
Algorithm~GQME satisfies Mutual Exclusion.
\begin{proof}
Suppose for contradiction that there exists an execution history $H \in \Hists$
at the end of which distinct processes $p$ and $q$ are both in the CS,
in passages $i$ and $j$, respectively.  
Let $\L{H} \in \Lin(H)$.
Without loss of generality,
  suppose that in $\L{H}$, $p$ executes $M.\enqueue$ in passage $i$
before $q$ executes $M.\enqueue$ in passage $j$.
Note that at the end of $\L{H}$, $p$ and $q$ are both in the CS,
    in passages $i$ and $j$, respectively,
    and in particular $p$ has not executed $M.\dequeue$ in passage $i$
    (since $p$ has not invoked $M.\dequeue$ in passage $i$ in $H$).
Thus, $\L{H}$, $p$, and $q$ contradict Lemma~\ref{i0_lem_fcfs}.
\end{proof}
\end{corollary}

\newpage
\begin{corollary} \label{i0_fcfs}
Algorithm~GQME satisfies First-Come First-Served.
\begin{proof} 
Suppose for contradiction that there exists an execution history $H \in \Hists$
in which process $p$ completes its execution of $M.\enqueue$ in passage $i$ before $q$
begins its execution of $M.\enqueue$ in passage $j$, and at the end of which $q$
is in the CS in passage $j$ but $p$ has not completed
the CS in passage $i$.
In particular, $p$ has not invoked $M.\dequeue$ in passage $i$ in $H$.    
Let $\L{H} \in \Lin(H)$.
Then $p$ executes $M.\enqueue$ in passage $i$ before $q$
    executes $M.\enqueue$ in passage $j$ in $\L{H}$.
Furthermore, at the end of $\L{H}$, $q$ is in the CS in passage $j$
    but $p$ has not executed $M.\dequeue$ in passage $i$
    (since $p$ has not invoked $M.\dequeue$ in passage $i$ in $H$).
Thus, $\L{H}$, $p$, and $q$ contradict Lemma~\ref{i0_lem_fcfs}.
\end{proof}
\end{corollary}

\begin{theorem} \label{i0_lf}
Algorithm~GQME satisfies Lockout Freedom.
\begin{proof} Suppose for contradiction that there is an infinite fair history $H \in \Hists$ 
in which some process $p$ begins some passage $i$ and then takes infinitely many 
   steps but never completes passage $i$.
By the structure of the algorithm, $p$ in passage $i$ loops forever
   at line~\ref{gm_4}, repeatedly reading $\Wayte[p] = \true$.
Let $E$ be a prefix of $H$ up to but not including the last step $(\ATM, p, M, \enqueue, \OK)$
(see line~\ref{gm_2}).
Choose $p$ so that $|E|$ is minimal.
Let $F$ be a prefix of $H$ up to and including the last step
$\left<(\ATM, p, M, \isHead, \rt)\right>$ for some response $\rt$.
Since $p$ loops forever at line~\ref{gm_4} it follows that $F$ exists, $E \preceq F$,
and $\rt = \false$.  Furthermore, $M^F \neq \bot$ by Theorem~\ref{i0_t1},
and $p \neq head(M^F)$ since $\rt = \false$, so $pred(M^F, p) \neq \bot$. 
Let $q = pred(M^F, p)$ and note that since $|E|$ minimal and since $H$ is fair,
$q$ eventually enters phase NEAR\_NCS in $H$ after $F$.
In particular, $q$ eventually executes 
$(\ATM, q, M, \dequeue, p)$, $(\ATM, q, \Wayte[p], write(\false), \OK)$, in that order, corresponding to line~\ref{gm_7} and line~\ref{gm_9}.
Now let $G$ be any prefix of $H$ such that $F \preceq G \preceq H$,
in which $q$ has executed the above two steps.
It follows that $\Wayte[p]^G = \false$, which contradicts $p$ 
   repeatedly reading $\Wayte[p] = \true$ at line~\ref{gm_4}
   in $H$ after the prefix $F$.
(We do not consider the possibility of $p$ looping forever during a $\dequeue$
   operation on $M$ because we assume in this section that $M$
   is an atomic base object.
 Later on we will show for each implementation of \MQ\
   that each operation on the implemented object incurs $\O(1)$ steps.)
\end{proof}
\end{theorem}

\begin{theorem}  \label{i0_be}
Algorithm~GQME satisfies the bounded exit property.
\begin{proof}
The result follows directly from the structure of Algorithm~GQME.
(We do not consider the number of steps incurred during a $\dequeue$
   operation on $M$ because we assume in this section that $M$
   is an atomic base object.
 Later on we will show for each implementation of \MQ\
   that a call to $\dequeue$ incurs $\O(1)$ steps.)
\end{proof}
\end{theorem}

\begin{theorem} \label{i0_rmr}
Algorithm~GQME has RMR complexity $\O(1)$ per passage in both the CC and DSM models
   provided that each operation on $M$ incurs $\O(1)$ RMRs.
\begin{proof}
Note that each passage involves only three \MQ\ operations,
at most two atomic write operations, 
and an unbounded number of atomic read operations at line~\ref{gm_4}.
So, it suffices to show that a process performs $\O(1)$ remote memory references at line~\ref{gm_4}.
This is obvious in the DSM model since $\Wayte[p]$ is local to $p$, in which case a process incurs zero RMRs on
line~\ref{gm_4}. Now consider the CC model. Note that $p$ incurs at most one RMR at line~\ref{gm_4} before $\Wayte[p] = \true$ is local to
$p$ (if this ever occurs). Also, $p$ is the only process that can assign $\Wayte[p] = \true$, so a
subsequent cache miss 
implies that $p$ reads $\Wayte[p] = \false$. Thus, $p$ breaks out of the busy-wait loop 
at line~\ref{gm_4} after at most two RMRs in total.
\end{proof}
\end{theorem}

\vfill


\section{Wait-free Implementation of \MQ\ Using Fetch-and-Increment \label{sec_fi}}
The implementation of an $N$-process \MQ\ object described in Figure~\ref{fig_fi_algo} is based on the mutual
exclusion algorithm of T. Anderson \cite{tand:spin}, as modified by J. Anderson and Y.-J. Kim for
efficient operation in the DSM model (see footnote 7 in \cite{jand:phi2}).
It relies on a shared object supporting a fetch-and-increment ($\FI()$) operation,
which atomically increments a variable and returns its previous value.
We assume that this shared object can also be reset to an initial value, e.g., via a write.

Implementation~MQFI (Figure~\ref{fig_fi_algo}) explicitly maintains a queue of processes
using a pair of circular arrays.  
When a process enqueues itself, it obtains an index in the two arrays by 
atomically incrementing variable $Ctr$ at line 1. Thus, the set of processes
enqueued at a given time maps to a contiguous (modulo $N$) block of array indices.
The array $Proc$ stores the IDs of enqueued processes (that are visible),
and array $Stat$ tracks the index of the head element and the visibility
of each process. 
Roughly speaking, this is done as follows: when a process $p$ enqueues itself
after a predecessor $q$, it is assigned array index $i$, where
$Stat[i] = 0$.  This value of $Stat[i]$  indicates that $p$ is neither visible nor the head of the \MQ.
$Stat[i]$ later becomes 1 if either $p$ becomes visible or $q$ dequeues itself, making $p$ the head element.
$Stat[i]$ becomes 2 once both $p$ has become visible and $q$ has dequeued itself.
Finally, $Stat[i]$ is reset back to 0 when $p$ dequeues itself.
Elements of $Stat$ are updated atomically using fetch-and-increment to ensure
that processes performing concurrent $\isHead$ and $\dequeue$ operations receive
consistent views of the \MQ\ object
(recall the discussion of race conditions in the second paragraph of Section~\ref{sec_contribs}).

\renewcommand{\tab}{\hspace{0.9cm}}
\begin{figure}\begin{center}
\begin{algorithm}[H]
\begin{tabbing}
Shared variables:   \\
 \tab \= $Stat$: \hspace{0.3cm} \= array $[0 .. N-1]$ of integer $0 .. 2$ \\
 \> \> initially $Stat[i] = \twodefo{1}{i = 0}{0}$ \\
\> $Proc$: \> array $[0 .. N-1]$ of integer $0 .. N-1$, uninitialized \\
\> $Ctr$: \> integer, initially zero \\
\\
Static private (per-process) variables: \\
\> $index$: \> integer $0 .. N-1$, uninitialized
\end{tabbing}
 Procedure for operation $\enqueue$ by process $p$: \; \Indp\Indp
\lnl{fi_e1}  $index \gets Ctr.\FI() \fmod N$    \; 
\lnl{fi_e2}  \Return{\OK}   \;  
\BlankLine \; 
\BlankLine \;
 Procedure for operation $\isHead$ by process $p$:  \Indm\Indm \; \Indp\Indp
\lnl{fi_i1}   $Proc[index].\opwrite(p)$   \;
\lnl{fi_i2}   \Return{$Stat[index].\FI() = 1$}  \;
\BlankLine \; 
\BlankLine \;
 Procedure for operation $\dequeue$ by process $p$: \Indm\Indm \; \Indp\Indp
\lnl{fi_d1}    $Stat[index].\opwrite(0)$   \;
\lnl{fi_d2}    \eIf{$Stat[(index + 1) \fmod N].\FI() = 1$}{
\lnl{fi_d3}    \Return{$Proc[(index + 1) \fmod N].\opread()$} \; 
     	            } {
\lnl{fi_d4}  \Return{$-1$} \;
       }
\end{algorithm}
\caption{\label{fig_fi_algo} Implementation~MQFI ($N$-process \MQ\ implementation using Fetch-and-Increment).}
\end{center}\end{figure}


\newcommand{\IFI}{I_{\textrm{MQFI}}}
\subsection{Proof of Correctness \label{sec_fi_poc}}
We denote Implementation~MQFI (shown in Figure~\ref{fig_fi_algo})
  of type $\MQ$ formally as
    $\IFI = (\Procs, \Vars, \Hists)$ where $\Procs = \{0..N-1\}$ and $\Vars$ consists of:
    the base objects $\{Ctr$, $Stat[0..N-1]$, $Proc[0..N-1]\}$,
          denoted subsequently as the set $\Base$, and a target object $M$.
Each $H \in \Hists$ is a two-level execution history 
    where processes call the procedures $\enqueue$, $\isHead$ and $\dequeue$
    as explained in Section~\ref{sec_mcd}.
For each such procedure call,
    $H$ records an invocation step on $M$ for the corresponding operation and,
	if the procedure call terminates, a matching response step on $M$
        with a response equal to the value returned by the procedure call.
Similarly, $H$ contains an atomic step for each
    operation that a process applies to one of the base objects $\Base$.

Implementation $\IFI$ simulates a \MQ\ object that can be used
  in Algorithm~GQME (Figure~\ref{fig_genalg}) provided that
  it is linearizable with respect to the \MQ\ type,
  and that each call to an access procedure incurs $\O(1)$ steps.
The latter property follows easily from the structure of the access procedures
  and also implies $\O(1)$ RMR complexity.
Therefore, we focus at linearizability.
Specifically, we must show that for every $H \in \Hists$, $H|M$ is linearizable
  with respect to the \MQ\ type.
To that end, we will explicitly construct a candidate linearization $\L{H}$ of $H|M$,
   and prove that $M$ conforms to the \MQ\ type.
We will do this using an invariant that relates the
   state of the base objects to the ``linearized state'' of $M$,
   which is determined by our candidate linearization.

Given $H \in \Hists$, we construct $\L{H}$ as follows.
Recall that in Figure~\ref{fig_fi_algo},
	the access procedure for each operation of type \MQ\
	contains one or more accesses to base objects.
For each \MQ\ operation execution in $H$,
	we define one of these base object accesses as
	the \emph{linearization point} of that operation execution.
Intuitively (and as we will prove in Theorem~\ref{i1_t1}),
	the order of the linearization points
	determines the order in which
	the MutexQueue operations that contain them are linearized.
Specifically, the linearization point of 
\begin{itemize}
\item an $\enqueue$ operation execution
	is the base object step $Ctr.\FI()$ at line~\ref{fi_e1};
\item an $\isHead$ operation execution
	is the base object step $Stat[index].\FI()$ at line~\ref{fi_i2};
\item a $\dequeue$ operation execution
	in which $Stat[(index+1)\fmod N].\FI()$ at line~\ref{fi_d2}
	returns $1$
	is the base object step $Proc[(index+1)\fmod N].\opread()$
	at line~\ref{fi_d3};	 and
\item a $\dequeue$ operation execution
	in which $Stat[(index+1) \fmod N].\FI()$ at line~\ref{fi_d2}
	returns a value other than $1$
	is that base object step itself.
\end{itemize}
Note that the response of a MutexQueue operation execution
	is uniquely determined if its linearization point has occurred.
For $\enqueue$,
	the response is always $\OK$.
For $\isHead$, the response is $\true$ if and only if
	the linearization point's response is 1.
For $\dequeue$,
	the response is the response of the linearization point,
	if the $\FI()$ at line~\ref{fi_d2}
	returns 1;
	and $-1$, otherwise.

For any $H \in \Hists$, let $\L{H}$ denote the complete sequential history over $M$
defined below, based on the linearization points present in $H$:
\begin{itemize}
\item $\L{H}$ contains each operation execution invoked in ${H}|M$
      whose linearization point appears in ${H}$, with the response
      determined by this linearization point, and no other steps.
\item Operation executions in $\L{H}$ occur in the same order as the corresponding linearization points
      in ${H}$.
\end{itemize}

Note that, by definition, $\L{H}$ is a history over the target object $M$, so we
can use the notation $succ(M^\L{H}, p)$ and $pred(M^\L{H}, p)$ defined in Section~\ref{sec_sup_ds}.
We also make extensive use of the following notation:
$index^{H}_p$ is the last value read from $Ctr$ by $p$ in ${H}$, reduced mod~$N$,
or $\bot$ if ${H}|p|Ctr = \bg{}$.
Informally, $index^{H}_p$ denotes the value of the private variable $index_p$ at the end of ${H}$,
assuming that $index$ is updated atomically with the response of $Ctr.\FI()$ at line~\ref{fi_e1} of $\enqueue$.

\begin{observation} \label{o_i1_prefix}
For any $G,H \in \Hists$ such that ${G} \preceq {H}$, $\L{G} \preceq \L{H}$.
\end{observation}

Informally, the following lemma says that two processes currently in the queue
cannot be assigned the same array index.

\begin{lemma}  \label{i1_l1}
For any $H \in \Hists$ and for any $p,q \in \Procs$ suppose that
$\L{H} \in \Lin({H}|M)$,
$M^{\L{H}} \neq \bot$, $p \in QProcs(M^{\L{H}})$, $q \in QProcs(M^{\L{H}})$, and
$index^{H}_p = index^{H}_q$.  Then $p = q$.
\begin{proof}
Suppose for contradiction that $p \neq q$.
Without loss of generality, assume that $p$'s last \enqueue\ in $\L{H}$ precedes $q$'s last \enqueue.
Then $q$ is the $k$'th process enqueued after $p$ in $\L{H}$
for some $k = mn$ and some $m \geq 1$. 
Let $(Q, V) = M^{\L{H}}$.  
It follows that $|Q| \geq k+1$ (i.e., $Q$ contains at least $p$ and a chain of
$k$ successors up to and including $q$).
Since $m \geq 1$ it follows that $k + 1 > N$, so by the pigeonhole principle
$Q$ contains two instances of some element, which contradicts Observation~\ref{o_dupno} (b).
\end{proof}
\end{lemma}

Next, define a \emph{bad} \MQ\ operation execution as one that violates the access etiquette for \MQ.
More precisely, if $H \in \Hists$ then
a \MQ\ operation execution $oe$ by process $p$ in ${H}$ is
bad if and only if there exists a prefix ${G}$ of ${H}$ that
contains the invocation of $oe$ but not its linearization point, 
such that $\L{G} \in \Lin({G}|M)$, $M^\L{G} \neq \bot$, and
one of the following holds:
\begin{itemize}
\item $oe$ is $\enqueue$ and $p \in QProcs(M^\L{G})$
\item $oe$ is $\isHead$ and either $p \not\in QProcs(M^\L{G})$ or $p \in V\!isProcs(M^\L{G})$ 
\item $oe$ is $\dequeue$ and either $p \neq head(M^\L{G})$ or $p \not\in V\!isProcs(M^\L{G})$
\end{itemize}

\bigskip
The following theorem establishes the correctness of Implementation~MQFI.

\begin{theorem} \label{i1_t1}
For any $H \in \Hists$, $H|M$ is linearizable with respect to type \MQ.

\begin{proof}
We will prove by induction on $|H|$ the following claim:

\begin{quote}
If ${H}$ does not contain any bad operation executions then $\L{H} \in \Lin({H}|M)$,
     $M^\L{H} \neq \bot$, and the values of the elements of $Stat$ at the end of ${H}$ are as follows:

\begin{eqnarray*}
  Stat[i]^{H} = \left\{
    \begin{array}{lll}
      2 &\mbox{if} &\exists p\in\Procs : index^{H}_p = i \we p = head(M^{\L{H}}) \we p \in V\!isProcs(M^{\L{H}}) \\
        & &\we \ \mbox{in ${H}$, $p$ has not written $Stat[i]$ at line~\ref{fi_d1} of $\dequeue$ since last}  \\
        & & \ \ \ \ \mbox{invoking $\enqueue$}  \\
      2 &\mbox{if} &\exists p\in\Procs : index^{H}_p = i \we p \neq head(M^{\L{H}}) \we p \in V\!isProcs(M^{\L{H}}) \\
        & &\we pred(M^\L{H}, p) \neq \bot \we pred(M^\L{H}, p) \mbox{ is between lines 6 and 7 } \\
        & & \ \ \ \mbox{of $\dequeue$ in ${H}$} \\
      1 &\mbox{if} &\exists p\in\Procs : index^{H}_p = i \we p \neq head(M^{\L{H}}) \we p \in V\!isProcs(M^{\L{H}}) \\
        & &\we (pred(M^\L{H}, p) = \bot \ve pred(M^\L{H}, p) \neq \bot \we pred(M^\L{H}, p) \mbox{ is not } \\
        & & \ \ \ \mbox{ between lines 6 and 7 of $\dequeue$ in ${H}$} ) \\
      1 &\mbox{if} &\exists p\in\Procs : index^{H}_p = i \we p = head(M^{\L{H}}) \we p \not\in V\!isProcs(M^{\L{H}}) \\
      1 &\mbox{if} & empty(M^{\L{H}}) \we i = Ctr^{{H}} \fmod N   \\
      0 & \multicolumn{2}{l}{\mbox{otherwise}}
   \end{array}
   \right.
\end{eqnarray*}
\end{quote}

    Informally, the above statement means the following.
    When $p$ enqueues itself, $Stat[index^{H}_p]$ is 1 if $p$ is the head of $M$ and 0 otherwise.
    $Stat[index^{H}_p]$ is subsequently incremented once when $p$ becomes visible, and once
    when $p$ becomes the head (or is about the become the head and its predecessor has partially completed $\dequeue$).
    The latter two operations may happen in either order.
    Finally, $Stat[index^{H}_p]$ returns to 0 once $p$ is visible, is the head of $M$ and has begun
    dequeuing itself (i.e., executed line~\ref{fi_d1}).  Furthermore, when $M$ is empty, $Stat[i] = 1$ if $i$
    is the array index that will be assigned to the next process that enqueues itself, and $Stat[i] = 0$ otherwise.

    Note that, by Lemma~\ref{i1_l1}, for every $i \in [0..N-1]$, 
    there is at most one $p \in QProcs(M^\L{H})$ such that $i = index^{H}_p$.  

In the remainder of the proof we denote the predicate that 
 $Stat[i]^{H}$ has the value specified above by $\invv({H}, i)$.

\medskip\noindent
\textbf{Basis:} $|H| = 0$.  It follows that ${H} = \L{H} = \bg{}$,
so certainly $\L{H} \in \Lin({H}|M)$.
Moreover, $empty(M^{\L{H}})$ holds,
so $\invv(H, i)$ follows from the initialization of Implementation~MQFI, for all $i \in [0..N-1]$.

\medskip\noindent
\textbf{Induction Hypothesis:} For any $l > 0$, assume
that Theorem~\ref{i1_t1} holds for every $H$ such that $|H| < l$.

\medskip\noindent
\textbf{Induction Step:} 
We must prove Theorem~\ref{i1_t1} for every ${H}$ such that $|H| = l$.
Let $G$ be a prefix of $H$ of length $l - 1$.
We proceed by cases on the last step $\sigma$ in ${H}$.
Cases A--G are when ${H}$ ends with an atomic base object step
and Case H is when ${H}$ ends with a non-atomic step on the target object $M$.
In all these cases we assume that ${H}$ does not contain a bad \MQ\ operation execution.
Finally, Case I is when ${H}$ does contain a bad \MQ\ operation execution.

\medskip\noindent
\textbf{Case A:} step $\sigma$ is a $Ctr.\FI()$
(see line~\ref{fi_e1} of \enqueue)
In this case, 
  \[ \L{H} = \L{G} \circ \bg{(\INV, p, M, \enqueue), (\RES, p, M, \OK)} \]
and $p \not\in QProcs(M^\L{G})$, since ${H}$ does not contain a bad operation execution.
Then certainly $\L{H} \in \Lin({H}|M)$, and $M^\L{H} \neq \bot$.
Furthermore, $p \in QProcs(M^\L{H})$, and $p \not\in V\!isProcs(M^\L{H})$.
Next, note that $Ctr^{H} = Ctr^{{G}} + 1$ and $Stat[i]^{H} = Stat[i]^{G}$
for all $i \in [0..N-1]$.  Let $j = index^{H}_p$ (i.e., $j = Ctr^{{G}} \fmod N$).
It remains to show $\invv({H}, i)$ for all $i \in [0..N-1]$.
For $i \neq j$ it follows from the IH that $Stat[i]^{H}$ has the value stipulated by
$\invv({H}, i)$.  Finally, consider $Stat[j]^{H}$.
\\
\textbf{Subcase A-i:} $empty(M^{\L{G}})$.  Then $p = head(M^{\L{H}})$
and $p \not\in V\!isProcs(M^{\L{H}})$, so we must show that $Stat[j]^{H} = 1$
(see fourth clause in the definition of $Stat[j]^{H}$).
But this follows from $Stat[j]^{H} = Stat[j]^{G}$ and $\invv({G}, j)$ (fifth clause), as wanted.
\\
\textbf{Subcase A-ii:} $\neg empty(M^{\L{G}})$.  Then $p \neq head(M^{\L{H}})$ 
and $p \not\in V\!isProcs(M^{\L{H}})$, so we must show that $Stat[j]^{H} = 0$
(see fifth clause in definition of $Stat[j]^{H}$).
By Lemma~\ref{i1_l1}, there is no $q \in QProcs(M^{\L{G}})$ such
that $q \neq p$ and $index^{H}_q = j$, so $Stat[j]^{H} = 0$ follows
from $Stat[j]^{H} = Stat[j]^{G}$ and $\invv({G}, j)$ (sixth clause), as wanted.

\medskip\noindent
\textbf{Case B:} step $\sigma$ is a
$Proc[index^{G}_p].\opwrite(p)$ (see line~\ref{fi_i1} of $\isHead$).
In this case, $\L{H} = \L{G}$, so $M^{\L{H}} = M^{\L{G}}$ and $M^{\L{H}} \neq \bot$ since
$M^{\L{G}} \neq \bot$ by the IH.
Furthermore, ${G}|M = {H}|M$, so $\L{H} \in \Lin({H}|M)$
since $\L{G} \in \Lin({G}|M)$ by the IH.

\medskip\noindent
\textbf{Case C:} step $\sigma$ is a
$Stat[index^{G}_p].\FI()$ with response $r$ for some $r$ (see line~\ref{fi_i2} of $\isHead$).
In this case,
    \[ \L{H} = \L{G} \circ \bg{(\INV, p, M, \isHead), (\RES, p, M, ret)} \]
where $ret = \true$ if $r = 1$ and $ret = \false$ otherwise.
Furthermore, $p \in QProcs(M^\L{G})$ and $p \not\in V\!isProcs(M^\L{G})$ since
${H}$ does not contain a bad operation execution.
Let $j = index^{H}_p$.  Then by $\invv({G}, j)$,
$r = 1$ if $p = head(M^{\L{G}})$ and $r = 0$ otherwise,
so it follows that $\L{H} \in \Lin({H}|M)$ and $M^\L{H} \neq \bot$.
Furthermore, $p \in QProcs(M^\L{H})$ and $p \in V\!isProcs(M^\L{H})$ hold.
It remains to prove $\invv({H}, i)$ for $i \in [0..N-1]$. 
For $i \neq j$, we have $Stat[i]^{H} = Stat[i]^{G}$, and 
$\invv({G}, i)$ implies $\invv({H}, i)$.
Finally, consider $Stat[j]^{H}$.  Note that $Stat[j]^{H} = Stat[j]^{G} + 1$, by 
the effect of the operation under consideration in this case.
\\
\textbf{Subcase C-i:} $p = head(M^{\L{G}})$.  Then $p = head(M^{\L{H}})$, and we must
show $Stat[j]^{H} = 2$ since $p \in V\!isProcs(M^\L{H})$
(see first clause in definition of $Stat[j]^{H}$),
i.e., we must show that $Stat[j]^{G} = 1$. But this follows from $\invv({G}, j)$ (fourth clause).
\\
\textbf{Subcase C-ii:} $p \neq head(M^{\L{G}})$.  Then $p \neq head(M^{\L{H}})$,
and we must show that $Stat[j]^{H} \in \bc{1, 2}$ since $p \in V\!isProcs(M^\L{H})$,
(see second and third clause in the definition of $Stat[j]^{H}$).
Since $p \neq head(M^{\L{G}})$, $p \in QProcs(M^\L{G})$ (so $M^\L{G}$ is not empty),
and $p \not\in V\!isProcs(M^\L{G})$,
$\invv({G}, j)$ implies that $Stat[j]^{G} = 0$, hence $Stat[j]^{H} = 1$, as wanted.

\medskip\noindent
\textbf{Case D:} step $\sigma$ is a
$Stat[index^{G}_p].\opwrite(0)$ (see line~\ref{fi_d1} of $\dequeue$).
In this case, $\L{H} = \L{G}$; thus $\L{H} \in \Lin({H}|M)$ (since, by the IH,
$\L{G} = \L{H}$ is a linearization of ${G}|M = {H}|M$), and
$M^{\L{H}} \neq \bot$ (since $M^{\L{G}} \neq \bot$ by the IH).
Furthermore, $p \in QProcs(M^\L{G})$, $p \in V\!isProcs(M^\L{G})$ and $p = head(M^\L{G})$
since ${H}$ does not contain a bad operation execution, hence
$p \in QProcs(M^\L{H})$, $p \in V\!isProcs(M^\L{H})$ and $p = head(M^\L{H})$.
It remains to prove $\invv({H}, i)$ for all $i \in [0..N-1]$
Let $j = index^{H}_p$ and note that for all $i \in [0..N-1]$, $i \neq j$,
$\invv({H}, i)$ follows directly from $\invv({G}, i)$.
Finally, $\invv({H}, j)$ holds since $p = head(M^{\L{H}})$, 
$p \in V\!isProcs(M^{\L{H}})$, and $Stat[j]^{H} = 0$ by the effect of
  step $\sigma$.
(See the sixth clause in the definition of $Stat[j]^{H}$, noting that, at the end of ${H}$,
$p$ has just completed line~\ref{fi_d1}.) 

\medskip\noindent
\textbf{Case E:}
step $\sigma$ is a $Stat[(index^{G}_p + 1) \fmod N].\FI()$
that returns $ret \neq 1$ (see line~\ref{fi_d2} of $\dequeue$). 
In this case,
    \[ \L{H} = \L{G} \circ \bg{(\INV, p, M, \dequeue), (\RES, p, M, -1)}. \]
Since, by assumption, ${H}$ contains no bad operation executions, $p \in QProcs(M^\L{G})$,
$p \in V\!isProcs(M^\L{G})$ and $p = head(M^\L{G})$.
By the IH, $M^\L{G} \neq \bot$ and so $M^\L{H} \neq \bot$.
Let $j = index^{H}_p$, $k = j + 1 \fmod N$, and $q = succ(M^\L{G}, p)$.
Thus, if $q \neq \bot$ then $q = head(M^\L{H})$.
Furthermore, we claim that if $q \neq \bot$ then $q \not\in V\!isProcs(M^\L{G})$.
For, if not, $\invv({G}, k)$ (third clause) would imply that $Stat[k]^{G} = 1$,
which would contradict the hypothesis of the case -- specifically that $ret \neq 1$.
Recall from the transition function of \MQ\ that a $\dequeue$ operation applied to a state
in which the head of the queue has no successor or has a successor that is not visible
returns $-1$.  Thus, $\L{H} \in \Lin({H}|M)$, as wanted.
It remains to show that  $\invv({H}, i)$ holds for all $i \in [0..N-1]$.
This follows immediately by the IH $\invv({G}, i)$ for all $i \neq j,k$.

To see that $\invv({H}, j)$ holds, we must prove that $Stat[j]^{H} = 0$.
(This is because $j = index^{H}_p$, $p \neq head(M^\L{H})$ and $p \not\in V\!isProcs(M^\L{H})$,
so clause six applies in the definition of $Stat[j]^{H}$.
We assume here that $N > 1$, so if $empty(M^\L{H})$ then $j \neq Ctr^{H} \fmod N$
since $k = Ctr^{H} \fmod N$ and $j \neq k$.  The case $N = 1$ is easy to show, noting that $j = k$.)
Since $Stat[j]^{H} = Stat[j]^{G}$, it suffices to prove that $Stat[j]^{G} = 0$.
Observing that $j = index^{G}_p$, $p = head(M^\L{G})$, $p \in V\!isProcs(M^\L{G})$ and
in ${G}$, $p$ has executed line~\ref{fi_d1} of $\dequeue$ since its last invocation of $\enqueue$, 
we conclude (see clause six in the definition of $Stat[j]^{G}$) that, $Stat[j]^{G} = 0$, as wanted.

Finally, to see that $\invv({H}, k)$ holds, we consider two cases.

\noindent\textbf{Subcase E-i:} $q = \bot$.  In this case, $empty(M^\L{H})$ and $k = Ctr^{H} \fmod N$.
Thus, we must prove that $Stat[k]^{H} = 1$ (see fifth clause in the definition of $Stat[k]^{H}$).
By the IH, $Stat[k]^{G} = 0$ (see sixth clause in the definition of $Stat[k]^{G}$).
By the effect of step $\sigma$,
$Stat[k]^{H} = Stat[k]^{G} + 1$.  Thus, $Stat[k]^{H} = 1$, as wanted.

\noindent\textbf{Subcase E-ii:} $q \neq \bot$.  As argued above, in this case $q \not\in V\!isProcs(M^\L{G})$,
hence $q \not\in V\!isProcs(M^\L{H})$.
Furthermore, $q = head(M^\L{H})$. 
Thus, we must prove that $Stat[k]^{H} = 1$ (see fourth clause in the definition of
$Stat[k]^{H}$).  We also have $q \neq head(M^\L{G})$ (because $p = head(M^\L{G})$ and $q$,
$p$'s successor, cannot be the same as $p$ by Observation~\ref{o_dupno} (b)).
Thus, by the IH, $Stat[k]^{G} = 0$ (see sixth clause in the definition of $Stat[k]^{G}$). 
By the effect of step $\sigma$,
$Stat[k]^{H} = Stat[k]^{G} + 1$.  Thus, $Stat[k]^{H} = 1$, as wanted.

\medskip\noindent
\textbf{Case F:}
step $\sigma$ is a
$Stat[(index^{G}_p + 1) \fmod N].\FI()$ with return value 1 (see line~\ref{fi_d2} of $\dequeue$). 
In this case, $\L{H} = \L{G}$; thus $\L{H} \in \Lin({H}|M)$ (since, by the IH,
$\L{G} = \L{H}$ is a linearization of ${G}|M = {H}|M$), and
$M^{\L{H}} \neq \bot$ (since $M^{\L{G}} \neq \bot$ by the IH).
Since ${H}$ does not contain a bad operation execution, $p \in QProcs(M^\L{G})$, $p \in V\!isProcs(M^\L{G})$
and $p = head(M^\L{G})$.
Let $j = index^{H}_p$, $k = j + 1 \fmod N$, and $q = succ(M^\L{G}, p)$.
It remains to prove that $\invv({H}, i)$ holds for all $i \in [0..N-1]$.
This follows immediately by the IH $\invv({G}, i)$ for all $i \neq j,k$.
The argument proving that $\invv({H}, j)$ holds is exactly as in Case E.
Finally, consider $\invv({H}, k)$.  Since $k = index^{G}_p$, $q = succ(M^\L{G}, p)$,
$q \neq head(M^\L{G})$ (by Observation~\ref{o_dupno} (b)), and 
$Stat[k]^{G} = 1$ by the hypothesis of this case, it follows by the IH $\invv({G}, k)$ that
$q \in V\!isProcs(M^\L{G})$ (see clause three of the definition of $Stat[k]^{G}$).
Furthermore, since ${H}$ does not contain any bad operation executions by the IH,
$q$ is not executing a pending $\dequeue$ in ${G}$, and has not yet reached line~\ref{fi_d1} since
last invoking $\enqueue$.
Thus, $k = index^{H}_p$, $q \in V\!isProcs(M^\L{H})$, $Stat[k]^{H} = Stat[k]^{G} + 1 = 2$ and
$q = head(M^\L{H})$ by the effect of step $\sigma$,
so $\invv({H}, k)$ holds (see first clause in the definition of $Stat[k]^{H}$).

\medskip\noindent
\textbf{Case G:} step $\sigma$ is a
$Proc[(index^{G}_p + 1) \fmod N].\opread()$ that returns $ret$ for some $ret$
(see line~\ref{fi_d3} of $\dequeue$). In this case, 
    \[ \L{H} = \L{G} \circ \bg{(\INV, p, M, \dequeue), (\RES, p, M, ret)}. \]
Let $j = index^{H}_p$, $k = j + 1 \fmod N$, and $q = succ(M^\L{G}, p)$.
Note that $p \in QProcs(M^\L{G})$, $p \in V\!isProcs(M^\L{G})$ and $p = head(M^\L{G})$
as in Case E, so $\L{H} \in \Lin({H}|M)$ provided that $ret = q$.
Also note that $q \neq \bot$, since if ${F} \preceq {G}$ where ${F}$ ends just before
$p$'s last $\FI()$ operation (i.e., line~\ref{fi_d2} of $\dequeue$) then $succ(M^\L{F}, p) \neq \bot$
follows from the arguments in Case F, and $succ(M^\L{F}, p) = succ(M^\L{G}, p)$.
Similarly, it follows that $q \in V\!isProcs(M^\L{G})$ and that $q$ has not begun executing $\dequeue$
by the end of ${G}$.
From Lemma~\ref{i1_l1} and Implementation~MQFI, it follows that no process has overwritten 
$Proc[k]$ since $q$ last wrote it, so $Proc[k] = q$, and $ret = q$,
which implies that $\L{H} \in \Lin({H}|M)$, as wanted.
Now, $\invv({H}, i)$ for $i \in [0..N-1]$, $i \neq k$
follows directly from $\invv({G}, i)$.
Finally, $\invv({G}, k)$ implies that $Stat[k]^{G} = 2$
since $q \neq head(M^\L{G})$, $q \in V\!isProcs(M^\L{G})$, and $pred(\L{H}, q) = p$
is between lines 6 and 7 at the end of ${G}$
(see second clause in definition of $Stat[k]^{G}$).
Since $Stat[k]^{H} = Stat[k]^{G} = 2$,  $q = head(M^\L{H})$, $q \in V\!isProcs(M^\L{H})$,
and $q$ has not started $\dequeue$ by the end of ${H}$, it follows that
$\invv({H}, k)$ holds
(see first clause in definition of $Stat[k]^{H}$).

\medskip\noindent
\textbf{Case H:} step $\sigma$ is a non-atomic step on the target object $M$ by process $p$.
\\
\textbf{Subcase H-i:} $\sigma$ is an invocation step.  Then $\L{H} = \L{G}$ by definition
since the linearization point of every \MQ\ operation occurs after the initial
invocation step.
Furthermore, $\L{G} \in \Lin({H}|M)$ since
$\L{G} \in \Lin({G}|M)$ and ${H} = {G} \circ \bg{s_I}$ where $s_I$ is an invocation.
Thus, $\L{H} \in \Lin({H}|M)$, and
$M^\L{H} \neq \bot$ since $M^\L{G} \neq \bot$ by the IH.
\\
\textbf{Subcase H-ii:} $\sigma$ is a response step.  Then the linearization point
of the operation execution corresponding to $\sigma$ has occurred in ${G}$, and so $\L{G}$
contains this operation execution.
Since $\L{G} \in \Lin({G}|M)$ by the IH,
it follows that $\L{G} \in \Lin({H}|M)$ provided that $\sigma$ and the
last step in $\L{H}|p$ have equal return values.
But the latter follows from our construction of $\L{H}$.
(Recall that for an operation execution that is pending in $H$,
   if the linearization point has occurred then the operation execution
   is completed with a matching response step in $\L{H}$
   that returns the uniquely-determined return value of the access procedure.)
Similarly, it follows that $\L{H} = \L{G}$.
Thus, $\L{H} \in \Lin({H}|M)$ and $M^\L{H} \neq \bot$ since
$\L{G} \in \Lin({G}|M)$ and $M^\L{G} \neq \bot$ by the IH.

\medskip\noindent
\textbf{Case I:} ${H}$ contains a bad \MQ\ operation.
Let $F$ be the prefix of ${H}$ up to but not including the first
invocation step $\sigma_I$ of a bad \MQ\ operation execution.
By the IH, $\L{F} \in \Lin(F|M)$ and $M^\L{F} \neq \bot$.
To obtain a linearization of ${H}|M$, first let $L = \L{G} \circ \bg{\sigma_I, \sigma_R}$
where $\sigma_R$ is a response matching $\sigma_I$, with an arbitrary return value.
Since $\sigma_I$ corresponds to a bad operation execution, it follows that
$L \in \Lin(({G} \circ \bg{\sigma_I})|M)$, and that $M^L = \bot$.
Finally, form $L'$ by appending to $L$ a complete operation execution on $M$
   for all remaining operation executions in ${H}|M$
(i.e., those that have been invoked but are not present in $L$),
say in the order of their invocation steps in ${H}$.
Once again assign the return value for each such operation execution arbitrarily.
Since $M^L = \bot$, it follows that $L' \in \Lin({H}|M)$.
\end{proof}
\end{theorem}

\subsubsection{RMR Complexity}
Each access procedure of Implementation~MQFI performs $\O(1)$ steps since there are no loops.  In particular, the RMR complexity of each access procedure is $\O(1)$.

\subsubsection{Bounded Memory Implementation}
A drawback of the above implementation is that $Ctr$ grows without bound.
We now discuss how to implement $Ctr$ using bounded memory. One approach, used by \cite{jand:surv},
is to atomically subtract $N$ from $Ctr$ whenever $N - 1$ is fetched from the $\FI()$ at line~\ref{fi_e2} of $\enqueue$.
This ensures that $Ctr$ never grows beyond $2N - 1$ (since at most $N - 1$ other processes can increment
$Ctr$ before $N$ is subtracted).  The drawback of this solution is that a fetch-and-add primitive
is needed in addition to (or in place of) fetch-and-increment.
Another solution, brought to our attention by Prasad Jayanti, is to allow $Ctr$ to overflow, provided that it
returns to zero without halting the execution. In particular, if $Ctr$ is an unsigned $m$-bit
integer and $N$ divides $2^m$, then it is easy to see that Implementation~MQFI remains correct
(i.e., the values assigned to $index$ are as before).


\renewcommand{\Proc}{{\Huge XXX-BUG-BUG-BUG-BUG-BUG-XXX}}

\newpage
\section{Wait-free Implementation of \MQ\ Using Fetch-and-Store \label{sec_fs}}
The implementation of an $N$-process \MQ\ presented in Figure~\ref{fig_fs_algo}
is based on the mutual exclusion algorithm of Craig \cite{craig:queue, craig:qtr}, 
in particular a variant brought to our attention by Prasad Jayanti.
It relies on a shared object supporting a fetch-and-store ($\Sw$) operation,
which atomically writes a variable and returns its previous value.
Without loss of generality, we assume that such
an object also supports an ordinary write operation.
(One can always simulate a write by applying a $\Sw$ and ignoring the response.)

Informally, Implementation~MQFS (Figure~\ref{fig_fs_algo}) works as follows.
At each point in time each process $p$ ``owns'' exclusively
	an index $\myIndex_p$ of array $\Queue$;
	the index owned by $p$ changes each time
	the process dequeues itself (see line~\ref{fs_d5}).
For this reason $\Queue$ has $N+1$ entries;
if a process dequeues itself at a time when all others are enqueued,
	it needs to acquire an index different from
	those owned by the other processes and from
	the index it previously owned.

The processes currently in the queue implicitly form a list,
	the first element of which is the head of the queue.
The shared variable $\Last$ contains the index
	owned by the last process in the queue.
(Whenever the queue is empty, $\Last$ contains an index
	not currently owned by any process.)
When process $p$ enqueues itself it uses $\Sw$ on $\Last$
	to find out its predecessor's index
	(which $p$ records in $\prevIndex_p$)
	and to atomically swap its own index into $\Last$
        (see line~\ref{fs_e2}).
The use of $\Sw$ to atomically read and update $\Last$
	ensures the integrity of the list of processes
	waiting in the queue;
it is not possible for two processes getting enqueued concurrently
	to consider the same process as their predecessor.

Recall from the specification of \MQ\ that
	the operation $\isHead$ has two objectives:
	(a)~to determine whether the process $p$ executing the operation
	is the head of the queue, and
	(b)~to make $p$ visible to its predecessor, 
	thereby ensuring that when the predecessor dequeues itself,
	it will ``wake up'' $p$.
In addressing the second objective we must contend with the possibility
	of $p$ becoming visible to its predecessor just as
	that predecessor is dequeueing itself.
This race condition is handled by appropriate use of $\Sw$.
We now explain how the implementation of \MQ\
	achieves these two objectives.

When process $p$ enqueues itself,
	it sets $\Queue[\myIndex_p]=(\myIndex_p, p)$
	(see line~\ref{fs_e1}).
When $p$ dequeues itself, 
	it sets $\Queue[\myIndex_p]$ to a value \emph{different from}
	$(\myIndex_p, \wild)$, specifically to $(\prevIndex_p, p)$
	(see line~\ref{fs_d1}).\footnote{
In this context ``$\wild$'' denotes a wildcard value.}
(We use $\Sw$ for this assignment because of
	the race condition mentioned above, as we will explain shortly.)

When it executes operation $\isHead$,
	process $p$ signals its predecessor that it has become visible
	by swapping the index it owns, $\myIndex_p$, and its ID,
	into the predecessor's position of array $\Queue$, namely
	$\Queue[\prevIndex_p]$;
	it records the old value of $\Queue[\prevIndex_p]$
	in $\tempIndex_p$ and $\tempId_p$ (see line~\ref{fs_i1}).
With this information, $p$ can determine if it is the head of the queue:
	this is the case if and only if its predecessor had dequeued itself
	by the time $p$ signalled that it is visible,
	i.e., if and only if $\tempIndex_p\ne\prevIndex_p$
	(see line~\ref{fs_i2}).

Finally, we explain how a process $p$ that is dequeuing itself
	ensures that it ``wakes up'' its successor,
	provided that the latter is visible.
As we have seen, when $p$ dequeues itself,
	it swaps $(\prevIndex_p, p)$ (where $\prevIndex_p\ne\myIndex_p$)
	into $\Queue[\myIndex_p]$,
	and records the old value of $\Queue[\myIndex_p]$
	into $\tempIndex_p$ and $\tempId_p$ (see line~\ref{fs_d1}).
There are two cases, depending on the value of $\tempIndex_p$.

\begin{enumerate}
\item
Process $p$ finds that $\tempIndex_p\ne\myIndex_p$.
In this case, $p$'s successor $q$ must have executed line~\ref{fs_i1}
	and swapped $(\myIndex_q, q)$
	(where $\myIndex_q\ne\myIndex_p$
	since no two processes can own the same index at the same time)
	into $\Queue[\prevIndex_q]$,
	i.e., into $\Queue[\myIndex_p$]
	(since $p$ is $q$'s predecessor).
This means that $q$ became visible before $p$ dequeued itself,
	and so $p$ is in charge of waking up $q$ when it is dequeued.
Indeed, in this case, $p$'s call to $\isHead$ returns $q$'s ID at line~\ref{fs_d3}.

\item
Process $p$ finds that $\tempIndex_p=\myIndex_p$.
In this case, $\Queue[\myIndex_p]$
	has not been changed by $p$'s successor
	since the time when $p$ enqueued itself
	and wrote $\myIndex_p$ into $\Queue[\myIndex_p]$
	(see line~\ref{fs_e1}).
This means that the successor of $p$ is not yet visible
	and so $p$ is not responsible for waking it up.
Accordingly, in this case $p$'s $\dequeue$ operation returns $-1$
	(see line~\ref{fs_d4}).
\end{enumerate}

%
%

\renewcommand{\tab}{\hspace{0.9cm}}
\begin{figure}\begin{center}
  \begin{algorithm}[H]
	\begin{tabbing}
	Shared variables:  \\
	\tab \= $\Queue$: \hspace{0.7cm}  \= array $[0..N]$ of integer $0..N$, 
        initially $\Queue[i] \neq i$ \\
  	 \> $\Last$: \> integer $0..N$, initially $N$ \\
	\\
	Static private (per-process) variables: \\	
 	\>  $\myIndex$: \> integer $0 .. N$, initially $p$ for process $p$ \\
 	\>  $\prevIndex$: \> integer $0 .. N$, uninitialized \\
	\>  $\tempIndex$: \> integer $0 .. N$, uninitialized \\
	\>  $\tempId$: \> integer $0 .. N-1$, uninitialized
	\end{tabbing}
 Procedure for operation $\enqueue$ by process $p$: \; \Indp\Indp
\lnl{fs_e1}  $\Queue[\myIndex].\opwrite((\myIndex, p))$  \;
\lnl{fs_e2}   $\prevIndex \gets \Last.\Sw(\myIndex)$  \;
\lnl{fs_e4}  \Return{\OK}  \;
	\BlankLine
 Procedure for operation $\isHead$ by process $p$:  \Indm\Indm \; \Indp\Indp
\lnl{fs_i1}     $(\tempIndex, \tempId) \gets \Queue[\prevIndex].\Sw((\myIndex, p))$ \;
\lnl{fs_i2}     \Return{$\tempIndex \neq \prevIndex$}  \;
	\BlankLine
 Procedure for operation $\dequeue$ by process $p$:  \Indm\Indm \; \Indp\Indp
\lnl{fs_d1}     $(\tempIndex, \tempId) \gets \Queue[\myIndex].\Sw((\prevIndex, p))$ \;
\lnl{fs_d2}     \eIf{$\tempIndex \neq \myIndex$}{ \;
\lnl{fs_d3}    $ret \gets \tempId$ \;
        }{ \;
\lnl{fs_d4}      $ret \gets -1$ \;
       } \;
\lnl{fs_d5}     $\myIndex \gets \prevIndex$ \;
\lnl{fs_d6}    \Return{$ret$}
  \end{algorithm}
\caption{\label{fig_fs_algo} Implementation~MQFS ($N$-process \MQ\ implementation using Fetch-and-Store).}
\end{center}

\end{figure}

\newcommand{\IFS}{I_{\textrm{MQFS}}}
\subsection{Proof of Correctness} \label{sec_fs_poc}
We proceed using the same approach as in Section~\ref{sec_fi}.
We denote Implementation~MQFS (shown in Figure~\ref{fig_fs_algo})
  of type $\MQ$ formally as
    $\IFS = (\Procs, \Vars, \Hists)$ where $\Procs = \{0..N-1\}$ and $\Vars$ consists of:
    the base objects $\{Last$, $\Queue[0..N-1]\}$, denoted subsequently as the set $\Base$, and
    a target object $M$.
Histories in $\Hists$ model the execution of Implementation~MQFS in a sense analogous
    to the one defined in Section~\ref{sec_fi_poc} for Implementation~MQFI.
As before, it follows easily that each call to an access procedure incurs $\O(1)$ steps,
    and so we focus on linearizability. 
To that end, we define for any $H \in \Hists$ a candidate linearization $\L{H}$
    using the same approach as in Section~\ref{sec_fi_poc}.
We also define bad operation executions exactly as in Section~\ref{sec_fi_poc}.
For the candidate linearization, we define the linearization point of
\begin{itemize}
\item an $\enqueue$ operation execution
	is the base object step $\Last.\Sw(\myIndex)$ at line~\ref{fs_e2}; and
\item an $\isHead$ operation execution
	is the base object step $\Queue[\prevIndex].\Sw$ at line~\ref{fs_i1}; and
\item a $\dequeue$ operation execution
        is the base object step $\Queue[\myIndex].\Sw$ at line~\ref{fs_d1}.
\end{itemize}

Note that as in Section~\ref{sec_fi_poc}, the response of a \MQ\ operation execution
	is determined uniquely if its linearization point has been reached.
For $\enqueue$,
	the response is always $\OK$.
For $\isHead$, the response is $\true$ if and only if
	the linearization point's response is different from the value
      of $\prevIndex$ for the calling process.
For $\dequeue$,
	the response is $-1$ if the $\Sw$ at line~\ref{fs_d1} returns
        an ordered pair of the form ($\myIndex, \wild$),
        and is the second element in this ordered pair otherwise.

In the proof of correctness of Implementation~MQFS it will be useful to refer to the values of
private variables at the end of histories in $\Hists$.
Let $H \in \Hists$ be a history such that $\L{H} \in \Lin(H|M)$\footnote{
$\Lin(H|M)$ is the set of linearizations of $H|M$, as defined in Section~\ref{sec_mcd}.
}
 and $M^\L{H} \neq \bot$.
Let $v_p$ be a private variable of process $p$ (i.e., one of $\myIndex_p$, $\prevIndex_p$
or $\tempIndex_p$).  We use $v_p^{H}$ to denote the value of $v_p$ at the end of ${H}$,
assuming that each assignment to a private variable of $p$ occurs at the same time as the
response of the last base object step by $p$ that precedes that assignment
in the execution corresponding to ${H}$.
Below we also use the notion of \emph{bad} operation executions, defined exactly as
in Section~\ref{sec_fi_poc}.

For any $H \in \Hists$, $p \in \Procs$ and $i \in [0..N]$, 
  we say that $p$ \emph{owns} $i$ at the end of ${H}$
  if and only if $\myIndex_p^{H} = i$.

We now state two observations in connection with the above definitions.
Informally, these say that:
\begin{enumerate}\renewcommand{\labelenumi}{(\alph{enumi})}	
	\item The value of $\myIndex_p$ after $p$ performs a $\dequeue$ operation
    is the value of $\prevIndex_p$ when $p$ performed the preceding
    $\enqueue$.  Intuitively, this is because of line~\ref{fs_d5}.
    \item The value of $\prevIndex_p$ after $p$ has enqueued itself
    is the value that $\myIndex_q$ had when $q$ was last in the queue, where
    $q$ is the processes that entered the queue just before $p$.
    Intuitively, this is because of line~\ref{fs_e2}.
    \end{enumerate}

\bigskip\noindent
More formally, we have:

\begin{observation}\label{fs_obs}
Let $H \in \Hists$ be a history where $\L{H} \in \Lin(H|M)$ and $M^\L{H} \neq \bot$,
and let ${G} \preceq {H}$ (note that $\L{G} \preceq \L{H}$).
\begin{enumerate}\renewcommand{\labelenumi}{(\alph{enumi})}	
	\item Let $p$ be any process such that $p \in QProcs(M^\L{G})$ and  
    $p$ executes $\dequeue$ exactly once in $\L{H}$ after $\L{G}$.
    Then $\prevIndex_p^{G} = \myIndex_p^{H}$.
    \item Let $p,q$ be any processes such that $q \in QProcs(M^\L{G})$, $p \in QProcs(M^\L{H})$,
    $q$ is the process that executes the last $\enqueue$ preceding the last $\enqueue$ of $p$ in $\L{H}$,
    and $q$ executes $\dequeue$ at most once in $\L{H}$ following $\L{G}$.
    Then $\myIndex_q^{G} = \prevIndex_p^{H}$.
    \end{enumerate}
\end{observation}

\begin{lemma}\label{l_fsprops}
Let ${H} \in \Hists$ be a history where $\L{H} \in \Lin({H}|M)$ and $M^\L{H} \neq \bot$.
Then the following statements hold:
\begin{enumerate}\renewcommand{\labelenumi}{(\arabic{enumi})}	
	\item $\forall x,y \in \Procs$, $x \neq y \cimps \myIndex_x^{H} \neq \myIndex_y^{H}$
	\item $\forall x,y \in QProcs(M^\L{H})$, $x \neq y \cimps \prevIndex_x^{H} \neq \prevIndex_y^{H}$
	\item $\forall x \in \Procs$, $y \in QProcs(M^\L{H}),$\ if\ $\myIndex_x^{H} = \prevIndex_y^{H}$\
          then\ $y = succ(M^\L{H}, x)$
	\item $\forall x \in QProcs(M^\L{H})$, $\myIndex_x^{H} \neq \prevIndex_x^{H}$
	\end{enumerate}
\begin{proof}
We proceed by induction on $|{H}|$.  It suffices to prove (1)--(3) since (4) follows immediately from (3):
if $p \in QProcs(M^\L{H})$ and $\myIndex_p^{H} = \prevIndex_p^{H}$ then
(3) implies that $p = succ(M^\L{H}, p)$, which contradicts Observation~\ref{o_dupno}.

\medskip\noindent
\textbf{Basis:} $|{H}| = 0$.  It follows that $\L{H} = {H} = \bg{}$.
By initialization, $\myIndex_p^{H} = p$ and $p \not\in QProcs(M^\L{H})$ hold for every $p \in \Procs$,
and so (1)--(3) hold for ${H}$.

\medskip\noindent
\textbf{Induction Hypothesis:} For any $l > 0$, assume that Lemma~\ref{l_fsprops}
holds for all ${H}$ such that $|{H}| < l$.

\medskip\noindent
\textbf{Induction Step:} We must prove Lemma~\ref{l_fsprops} for every ${H}$ such that $|{H}| = l$.
Let $G$ be a prefix of $H$ of length $l - 1$.
We proceed by cases on the last step $\sigma$ in ${H}$.
Since $M^\L{H} \neq \bot$, it follows that $M^\L{G} \neq \bot$.

\medskip\noindent
\textbf{Case A:} $\L{G} = \L{H}$ or $\sigma$ is the linearization
point of $\isHead$ (line~\ref{fs_i1}).  In this case, for each $p \in \Procs$,
$\myIndex_p^{G} = \myIndex_p^{H}$ and $\prevIndex_p^{G} = \prevIndex_p^{H}$.
Moreover, $QProcs(M^\L{G}) = QProcs(M^\L{H})$.
Thus, the fact that the lemma holds for ${H}$ follows directly from the fact that
(by the IH) it holds for ${G}$.

\medskip\noindent
\textbf{Case B:} $\sigma$ is the linearization point of $M.\enqueue$ by process $p$.
Lemma~\ref{l_fsprops}~(1) for ${H}$ follows directly from the IH since
for every $x \in \Procs$, $\myIndex_x^{G} = \myIndex_x^{H}$.
It remains to prove parts (2) and (3) of the lemma for ${H}$.

\smallskip\noindent
\textbf{Subcase B1:} $p = head(M^\L{H})$.  It follows that $M^\L{G}$ is empty
and $QProcs(M^\L{H})$ contains only $p$, and so Lemma~\ref{l_fsprops}~(2) holds 
trivially for ${H}$.  Now let $j = \prevIndex_p^{H}$.
To prove part (3), it suffices to show that no process $z \in \Procs$ owns $j$
at the end of ${H}$.
Suppose for contradiction that for some $z \in \Procs$ $\myIndex_z^{H} = j$.
It follows that $\L{H}$ contains more than one $M.\enqueue$, otherwise $j = N$ and
$\myIndex_z^{H} = z$ where $z \neq N$.
Let $r$ be the process that executes the last $\enqueue$ preceding the
last $\enqueue$ of $p$ in $\L{H}$.
Let ${F}$ be the prefix of ${H}$ up to but not including the linearization point
  of the last $M.\dequeue$ performed by $r$; 
this is well-defined because $M^\L{G}$ is empty.
By Observation~\ref{fs_obs}~(b) and the fact that $j = \prevIndex_p^{H}$,
it follows that $\myIndex_r^{F} = j$.
Also note that no process other than $r$ applies a \MQ\ operation execution
  in $\L{G}$ after $\L{F}$.
There are two cases, each leading to a contradiction.
	\begin{itemize}
\item If $z \neq r$ then $\myIndex_z^{F} = j$ since $\myIndex_z^{H} = j$ and $z$
    does not execute $\dequeue$ in $\L{G}$ after $\L{F}$.   
    At the same time $\myIndex_r^{F} = j$, as argued above.
    But $\myIndex_z^{F} = j$ and $\myIndex_r^{F} = j$
    contradict part (1) of the IH for ${F}$ since $z \neq r$.
\item If $z = r$ then by Observation~\ref{fs_obs}~(a) and the fact that $\myIndex_z^{H} = j$,
    $\prevIndex_z^{F} = j$ and hence $\prevIndex_r^{F} = j$.
    At the same time, $\myIndex_r^{F} = j$, as argued above.  Furthermore,
    $r \in QProcs(M^\L{F})$ by definition of $r$ and ${F}$.
    But $\prevIndex_r^{F} = j$, $\myIndex_r^{F} = j$ and $r \in QProcs(M^\L{F})$
    contradict part (4) of the IH for ${F}$.
	\end{itemize}
Thus, Lemma~\ref{l_fsprops}~(3) holds for ${H}$.

\smallskip\noindent
\textbf{Subcase B2:} $p \neq head(M^\L{H})$. 
Let $r = pred(M^\L{H}, p)$ and let $j = \prevIndex_p^{H}$. 

First, consider Lemma~\ref{l_fsprops}~(2) for ${H}$.  
For every $q \in \Procs \setminus \bc{p}$, 
$\prevIndex_q^{G} = \prevIndex_q^{H}$ and
$q \in QProcs(M^{G}) \Leftrightarrow q \in QProcs(M^{H})$ hold, so it suffices
to show that there is no $z \in QProcs(M^\L{G})$ such that $\prevIndex_z^{G} = j$.
Suppose for contradiction that $\prevIndex_z^{G} = j$ for some $z \in QProcs(M^\L{G})$.
Observe that $r \in QProcs(M^\L{G})$ by the definition of ${G}$ and the
hypothesis of Subcase B2, and that $succ(M^\L{G}, r) = \bot$ by 
the definition of ${G}$ and the hypothesis of Case B.
Since $\prevIndex_p^{H} = j$, it follows from the definition
of $r$ and ${G}$ and Observation~\ref{fs_obs}~(b) that $\myIndex_r^{G} = j$.
Since $\prevIndex_z^{G} = j$ and $z \in QProcs(M^\L{G})$ by assumption, part (3) of the IH 
for ${G}$ implies that $z = succ(M^\L{G}, r)$.
But this contradicts the earlier observation that $succ(M^\L{G}, r) = \bot$.

Next, consider part (3) of the lemma.
It suffices to show that for any $q \in \Procs$, 
if $\myIndex_q^{H} = j$ then $p = succ(M^\L{H}, q)$.
By part (1) of the IH for ${G}$, $r$ is the only process 
that owns $\myIndex_r^{G}$ at the end of ${G}$, and so by the hypothesis of Case B,
$r$ is the only process that owns $\myIndex_r^{H}$ at the end of ${H}$.
By definition, $r = pred(M^\L{H}, p)$ and so $p = succ(M^\L{H}, r)$. 
Thus, Lemma~\ref{l_fsprops}~(3) holds for ${H}$.

\medskip\noindent
\textbf{Case C:} $\sigma$ is the
linearization point of $M.\dequeue$ by process $p$.
Note that $p = head(M^\L{G})$ since $M^\L{H} \neq \bot$.
Let $j = \prevIndex_p^{G}$.

First, consider Lemma~\ref{l_fsprops}~(1) for ${H}$.
Note that for every $q \in \Procs \setminus \bc{p}$,
$\myIndex_q^{G} = \myIndex_q^{H}$ holds.
Furthermore, $\myIndex_p^{H} = j$ by line~\ref{fs_d5} and Observation~\ref{fs_obs}~(a).  
It suffices to show that no process owns $j$ at the end of ${H}$.
Suppose for contradiction that $\myIndex_z^{H} = j$ for some $z \in \Procs \setminus \bc{p}$.
Then $\myIndex_z^{G} = j$, $\prevIndex_p^{G} = j$ and $p \in QProcs(M^\L{G})$
all hold by definition of ${G}$ and the hypothesis of Case C,
so by part (3) of the IH for ${G}$ it follows
that $p = succ(M^\L{G}, z)$.
But this contradicts the earlier observation that $p = head(M^\L{G})$.

Next, consider Lemma~\ref{l_fsprops}~(2) for ${H}$.
Note that for every $q \in \Procs \setminus \bc{p}$,
$\prevIndex_q^{G} = \prevIndex_q^{H}$ and
$q \in QProcs(M^{G}) \Leftrightarrow q \in QProcs(M^{H})$ hold.
Furthermore, $p \not\in QProcs(M^\L{H})$ by the hypothesis of Case C.
Thus, Lemma~\ref{l_fsprops}~(2) for ${H}$ follows directly from 
part (2) of the IH for ${G}$.

Finally, consider part (3).
Note that for every $q \in \Procs \setminus \bc{p}$, the following 
all hold:
$\myIndex_q^{G} = \myIndex_q^{H}$, 
$\prevIndex_q^{G} = \prevIndex_q^{H}$ and
$q \in QProcs(M^{G}) \Leftrightarrow q \in QProcs(M^{H})$.
Furthermore, $p \not\in QProcs(M^\L{H})$ by the hypothesis of Case~C.
Thus, by part (3) of the IH for ${G}$, it suffices to show that
there is no $z \in QProcs(M^\L{H})$ such that $\prevIndex_z^{H} = \myIndex_p^{H}$.
Suppose for contradiction that $\prevIndex_z^{H} = \myIndex_p^{H}$ for some $z \in QProcs(M^\L{H})$.
Note that $z \neq p$ since $p \not\in QProcs(M^\L{H})$,
and so by the hypothesis of Case C we further have
$\prevIndex_z^{G} = \prevIndex_z^{H}$ (hence $\prevIndex_z^{G} = \myIndex_p^{H}$)
and $z \in QProcs(M^\L{G})$.
At the same time, by line~\ref{fs_d5}, Observation~\ref{fs_obs}~(a) and the hypothesis of Case~C,
$\prevIndex_p^{G} = \myIndex_p^{H}$ and $p \in QProcs(M^\L{G})$ both hold.
Thus, we have shown that the following all hold:
$\prevIndex_z^{G} = \myIndex_p^{H}$, $\prevIndex_p^{G} = \myIndex_p^{H}$,
$z,p \in QProcs(M^\L{G})$ and $z \neq p$.  But this contradicts part (2) of the IH for ${G}$.
\end{proof}
\end{lemma}

\bigskip
The following theorem establishes the correctness of Implementation~MQFS.

\begin{theorem} \label{i2_t1}
For any $H \in \Hists$, $H|M$ is linearizable with respect to type \MQ.
\begin{proof}
We will prove by induction on $|H|$ the following claim:

\begin{quote}
    If ${H}$ does not contain any bad operation executions then $\L{H} \in \Lin({H}|M)$,
     $M^\L{H} \neq \bot$, and the value of $\Queue$ at the end of ${H}$
     is as follows:

    For any $i \in [0..N]$, if $\exists p\in \Procs$ such that at the end of ${H}$ $p$ owns index $i$,
	      and has applied $\Queue[\myIndex].\opwrite$ at line~\ref{fs_e1}
          of $\enqueue$, but since last doing so $p$ has not applied
          $\Queue[\myIndex].\Sw$ at line~\ref{fs_d1} of $\dequeue$, 
          then (letting $s$ denote $succ(M^\L{H}, p)$)
          \[ \Queue[i]^{H} = \twodefo{(i, p)}
            {s \not\in V\!isProcs(M^\L{H})}{(\myIndex^{H}_s,s)}
          \]
          else $Queue[i]^{H} \neq (i, \wild)$.
\end{quote}

    In the remainder of the proof we denote by $\invv({H}, i)$
    the predicate that at the end of execution history ${H}$, 
    $\Queue[i]^{H}$ has the value specified above.

\medskip\noindent
\textbf{Basis:} $f({H}) = 0$.  It follows that ${H} = \L{H} = \bg{}$,
so certainly $\L{H} \in \Lin({H}|M)$.
It remains to show $\invv(H, i)$ for $i \in [0..N]$, which in this case
  asserts that $\Queue[i]^{H} \neq (i, \wild)$.
But this follows from the initialization of Implementation~MQFS.

\medskip\noindent
\textbf{Induction Hypothesis:} For any $l > 0$, assume
that Theorem~\ref{i2_t1} holds for every $H$ such that $f({H}) < l$.

\medskip\noindent
\textbf{Induction Step:} 
We must prove Theorem~\ref{i2_t1} for every ${H}$ such that $|H| = l$.
Let $G$ be a prefix of $H$ of length $l - 1$.
We proceed by cases on the last step $\sigma$ in ${H}$.
Cases A--E are when ${H}$ ends with an base object step,
and Case F is when ${H}$ ends with a non-atomic step on the target object $M$.
In all these cases we assume that ${H}$ does not contain a bad \MQ\ operation execution.
Finally, Case G is when ${H}$ does contain a bad \MQ\ operation execution.


\medskip\noindent
\textbf{Case A:} $\sigma$ is a
$\Queue[\myIndex^{G}_p].\opwrite((\myIndex^{G}_p,p))$ (see line~\ref{fs_e1} of \enqueue). 
In this case, $\L{H} = \L{G}$; thus $\L{H} \in \Lin({H}|M)$ (since, by the IH,
$\L{G} = \L{H}$ is a linearization of ${G}|M = {H}|M$), and
$M^{\L{H}} \neq \bot$ (since $M^{\L{G}} \neq \bot$ by the IH).

To prove the theorem for ${H}$
it suffices to verify that $\invv({H}, \myIndex^{H}_p)$ holds;
all other clauses 
either hold trivially (because their antecedents are false) or follow immediately from the IH.
(Note that in this case $\myIndex_p^{H}$ is the only position in $\Queue$ changed
by $\sigma$, which does not change the linearized state of $M$.)
Since $p$ has just completed line~\ref{fs_e1} at the end of ${H}$, 
$\invv({H}, \myIndex^{H}_p)$ asserts that
$\Queue[\myIndex^{H}_p]^{H} = (\myIndex^{H}_p, p)$,
which indeed holds by the action of step $\sigma$.

\medskip\noindent
\textbf{Case B:} $\sigma$ is a $\Last.\Sw$
(see line~\ref{fs_e2} of \enqueue)
In this case, 
  \[ \L{H} = \L{G} \circ \bg{(\INV, p, M, \enqueue), (\RES, p, M, \OK)} \]
and $p \not\in QProcs(M^\L{G})$, since ${H}$ does not contain a bad operation execution.
Then certainly $\L{H} \in \Lin({H}|M)$, and $M^\L{H} \neq \bot$.

In the case under consideration, 
all clauses of Theorem~\ref{i2_t1} for ${H}$ 
either hold trivially (because their antecedents are false) or follow immediately from the IH.

\medskip\noindent
\textbf{Case C:} $\sigma$ is a
$\Queue[\prevIndex^{G}_p].\Sw$ with response $r$ for some $r$ (see line~\ref{fs_i1} of $\isHead$).
In this case,
    \[ \L{H} = \L{G} \circ \bg{(\INV, p, M, \isHead), (\RES, p, M, ret)} \]
where $ret = \true$ if $r \neq (\prevIndex_p^{H}, \wild)$ and $ret = \false$ otherwise.
Furthermore, $p \in QProcs(M^\L{G})$ and $p \not\in V\!isProcs(M^\L{G})$ since
${H}$ does not contain a bad operation execution.
To show that $\L{H} \in \Lin(H|M)$ we must show that
$ret = \true$ iff $p = head(M^\L{H})$.
Let $j = \prevIndex^{G}_p$ and consider the following subcases.

\smallskip\noindent
\textbf{Subcase C1:} Some $q \in \Procs$ owns $j$ at the end of ${G}$.
Then $p = succ(M^\L{G}, q)$ by Lemma~\ref{l_fsprops}~(3), and
in particular $q \in QProcs(M^\L{G})$ by definition of $succ$.
Furthermore $\Queue[j]^{G} = (j, q)$ by the IH for ${G}$ since
$p \not\in V\!isProcs(M^\L{G})$.
Thus, $r = (j,q)$ and so $ret = \false$, while $p \neq head(M^\L{G})$,
    hence $p \neq head(M^\L{H})$, as wanted.

To prove the theorem for ${H}$ in the case under consideration,
it suffices to verify that $\invv({H}, j)$ and $\invv({H}, myIndex^{H}_q)$ hold;
all other clauses of Theorem~\ref{i2_t1} for ${H}$ either hold trivially
or follow immediately from the IH.
(Note that in this case $j$ is the only position of $\Queue$ that is changed by $\sigma$, and $q$ is the only process whose successor, namely $p$, becomes visible
as a result of step $\sigma$; the linearized state of $M$ is otherwise unchanged.)
It follows by line~\ref{fs_e2} of the algorithm that $\myIndex^{H}_q = \prevIndex^{H}_p$,
and since $\prevIndex^{H}_p = \prevIndex^{G}_p$ that $\prevIndex^{H}_p = j$.
Thus, the conditions $\invv({H}, \myIndex^{H}_q)$ and $\invv({H}, j)$ are equivalent.
Furthermore, since $p \in V\!isProcs(M^\L{H})$ by the action of step $\sigma$,
$\invv({H}, \myIndex^{H}_q)$ asserts that $\Queue[\myIndex^{H}_q]^{H} = (\myIndex^{H}_p,p)$.
Indeed we have
    \[
    \begin{array}{rlll}
        \Queue[\myIndex^{H}_q]^{H} &= & \Queue[\prevIndex^{H}_p]^{H} & \mbox{because $myIndex^{H}_q = prevIndex^{H}_p$} \\
&= & \Queue[\prevIndex^{G}_p]^{H} & \mbox{because $\prevIndex^{H}_p = \prevIndex^{G}_p$} \\
    &=& (\myIndex^{G}_p,p) & \mbox{by the action of the last operation} \\ 
    &&& \mbox{execution in ${H}$} \\
    &=& (\myIndex^{H}_p,p) & \mbox{because $\myIndex^{H}_p = \myIndex^{G}_p$} 
    \end{array}
    \]

\smallskip\noindent
\textbf{Subcase C2:} No process owns $j$ at the end of ${G}$.
It follows that $p = head(M^\L{G})$, otherwise by line~\ref{fs_e2} 
and the fact that ${H}$ contains no bad operation executions it would be the case that
$\myIndex_{pred(M^\L{G}, p)}^{G} = j$.
Furthermore $\Queue[j]^{G} \neq (j,\wild)$ by the IH for ${G}$.
Thus, $r \neq (j, \wild)$ and so $ret = \true$, while $p = head(M^\L{G})$, as wanted.

To prove the theorem for ${H}$ in the case under consideration
it suffices to verify that $\invv({H}, j)$ holds;
all other clauses of Theorem~\ref{i2_t1} for ${H}$ either hold trivially
or follow immediately from the IH.
(Note that in this subcase the fact that $p$ becomes visible as a result of step $\sigma$ does not affect the value of 
$\Queue$ at the position owned by $p$'s predecessor, since $p = head(M^\L{G})$ and so $p$
has no predecessor.)
By the hypothesis of subcase C2, no process owns $j$ at the end of ${G}$,
hence no process owns $j$ at the end of ${H}$.
Thus, $\invv({H}, j)$ asserts that $\Queue[j]^{H} \neq (j, \wild)$.
Indeed we have
    \[
    \begin{array}{rlll}
        \Queue[j]^{H} &= &  (\myIndex^{G}_p,p) & \mbox{by the action of $\sigma$}\\
    &\neq& (\prevIndex^{G}_p,\wild) & \mbox{by Lemma~\ref{l_fsprops}~(4) for ${G}$}  \\
    &=& (j, \wild) 
    \end{array}
    \]

\medskip\noindent
\textbf{Case D:} $\sigma$ is a
$\Queue[\myIndex^{G}_p].\Sw$ with response $(\myIndex_p^{G}, \wild)$ (see line~\ref{fs_d1} of $\dequeue$). 
In this case,
    \[ \L{H} = \L{G} \circ \bg{(\INV, p, M, \dequeue), (\RES, p, M, -1)}. \]
Since, by assumption, ${H}$ contains no bad operation executions, $p \in QProcs(M^\L{G})$,
$p \in V\!isProcs(M^\L{G})$ and $p = head(M^\L{G})$.
By the IH, $M^\L{G} \neq \bot$ and so $M^\L{H} \neq \bot$.
By the case under consideration, $\Queue[\myIndex^{G}_p]^{G} = (\myIndex^{G}_p,\wild)$.
By the IH, $\invv({G}, \myIndex^{G}_p)$ holds, which implies along with
  Lemma~\ref{l_fsprops}~(1) that $succ(M^\L{G}, p) \not\in V\!isProcs(M^\L{G})$.
By the specification of \MQ\, the response of a $\dequeue$ operation by $p$ applied
to $M^\L{G}$ is $-1$. Thus, $\L{H} \in \Lin({H}|M)$.

To prove the theorem for ${H}$ in the case under consideration,
it suffices to verify that $\invv({H}, \myIndex^{G}_p)$ and $\invv({H}, \myIndex^{H}_p)$
both hold;
all other clauses for ${H}$ either hold trivially
or follow immediately from the IH.
(Recall our convention regarding when assignments to private variables take effect;
in particular, in this case, the assignment to $\myIndex_p$ at line~\ref{fs_d5} takes
effect atomically with step $\sigma$.  Thus, step $\sigma$
causes $p$ to relinquish ownership of $\myIndex_p^{G}$ and acquire ownership
of $\myIndex_p^{H}$.
Also, recall that $p$ has no predecessor in $M^\L{G}$, and so $p$ no longer
being visible has no impact on the meaning of $\invv({H}, i)$
for $i \neq \myIndex^{G}_p, \myIndex^{H}_p$.)

First consider $\invv({H}, \myIndex^{G}_p)$.  Notice that no process 
owns $\myIndex_p^{G}$ at the end of ${H}$.
This is because only $p$ owns $\myIndex_p^{G}$ at the end of ${G}$ (by Lemma~\ref{l_fsprops}~(1)),
and at the end of ${H}$ $p$ owns a different index, namely $\myIndex_p^{H}$ (note that
$\myIndex_p^{H} = \prevIndex_p^{G}$ by line~\ref{fs_d5}, and $\prevIndex_p^{G} \neq \myIndex_p^{G}$
by Lemma~\ref{l_fsprops}~(4)).  
Thus, $\invv({H}, \myIndex^{G}_p)$ asserts that $\Queue[\myIndex_p^{G}]^{H} \neq (\myIndex_p^{G},\wild)$.
This is indeed true since
    \[
    \begin{array}{rlll}
        \Queue[\myIndex_p^{G}]^{H} &= &  (\prevIndex^{G}_p,\wild) & \mbox{by the action of step $\sigma$} \\
    &\neq& (\myIndex^{G}_p, \wild) & \mbox{by Lemma~\ref{l_fsprops}~(4) for ${G}$}
    \end{array}
    \]

Next, consider $\invv({H}, \myIndex^{H}_p)$.
By the hypothesis of Case D, in ${H}$ $p$ has applied
$\Queue[\myIndex_p].\Sw$ at line~\ref{fs_d1} since it last applied
$\Queue[\myIndex_p].\opwrite$ at line~\ref{fs_e1}.  
Thus, $\invv({H}, \myIndex^{H}_p)$ asserts that
$\Queue[\myIndex_p^{H}]^{H} \neq (\myIndex_p^{H},\wild)$.
We now prove that this is the case.  We have that $\myIndex_p^{H} = \prevIndex_p^{G}$
(by line~\ref{fs_d5}). Also, no process owns $\prevIndex_p^{G}$ at the end of ${G}$:
$p$ does not own it because it owns $\myIndex_p^{G}$ and $\myIndex_p^{G} \neq \prevIndex_p^{G}$
(by Lemma~\ref{l_fsprops}~(4)); and no other process owns it
because if one, say $z$, did then $p$ and $z$ would both own it at the end of ${H}$,
contradicting Lemma~\ref{l_fsprops}~(1).  Since no process owns $\prevIndex_p^{G}$
at the end of ${G}$, by $\invv({G}, \prevIndex^{G}_p)$, which holds by the IH,
we have $\Queue[\prevIndex_p^{G}]^{G} \neq (\prevIndex_p^{G},\wild)$.
But then, since $\prevIndex_p^{G} = \myIndex_p^{H}$, and since step $\sigma$
does not change position $\prevIndex_p^{G}$ of $\Queue$ (it changes position
$\myIndex_p^{G}$, which is different from $\prevIndex_p^{G}$ by Lemma~\ref{l_fsprops}~(4)),
we have $\Queue[\myIndex_p^{H}]^{H} \neq \myIndex_p^{H}$, as wanted.

\medskip\noindent
\textbf{Case E:} $\sigma$ is a
$\Queue[\myIndex^{G}_p].\Sw$ with response different from $(\myIndex_p^{G},\wild)$ (see line~\ref{fs_d1} of $\dequeue$). 
In this case,
    \[ \L{H} = \L{G} \circ \bg{(\INV, p, M, \dequeue), (\RES, p, M, r)} \]
where $r = \tempId^{H}_p$
Since, by assumption, ${H}$ contains no bad operation executions, $p \in QProcs(M^\L{G})$,
$p \in V\!isProcs(M^\L{G})$ and $p = head(M^\L{G})$.
By the IH, $M^\L{G} \neq \bot$ and so $M^\L{H} \neq \bot$.
By the case under consideration, $\Queue[\myIndex^{G}_p]^{G} \ne (\myIndex^{G}_p,\wild)$.
By the IH, $\invv({G}, \myIndex^{G}_p)$ holds, which implies
  that $succ(M^\L{G}, p) \in V\!isProcs(M^\L{G})$ and
  moreover that $r = \tempId^{H}_p = succ(M^\L{G}, p)$.
By the specification of \MQ\, the response of a $\dequeue$ operation execution by $p$ applied
to $M^\L{G}$ is $succ(M^\L{G}, p)$.
Since $r = succ(M^\L{G}, p)$, $\L{H} \in \Lin({H}|M)$.

To prove the theorem for ${H}$ in the case under consideration,
  we proceed as in Case D.

\medskip\noindent
\textbf{Case F:} ${H}$ ends with an event on the target object $M$ by process $p$.
The proof is analogous to the one given in Case H in the proof of Theorem~\ref{i1_t1}
for Implementation~MQFI.

\medskip\noindent
\textbf{Case G:} ${H}$ contains a bad \MQ\ operation.
The proof is analogous to the one given in Case I in the proof of Theorem~\ref{i1_t1}
for Implementation~MQFI.
\end{proof}
\end{theorem}

\subsubsection{RMR Complexity}
Each access procedure of Implementation~MQFS incurs $\O(1)$ steps since there are no loops.  In particular, the RMR complexity of each access procedure is $\O(1)$.



\section{Conclusion \label{sec_conc}}
In this paper we have shown how to solve mutual exclusion for $N$ processes
    using a linearizable implementation of an $N$-process \MQ\ object
    and atomic read/write registers.
In doing so we have re-cast the problem of implementing and proving correct
    an $\O(1)$-RMR (per-passage) queue-based mutual exclusion algorithm
    into the intuitively more fundamental problem of implementing the underlying queue
    using $\O(1)$ RMRs per operation.
We have presented and proved correct two such implementations of \MQ,
    based on the mutual exclusion algorithms of T.\ Anderson and Craig \cite{tand:spin, craig:queue}.
We believe that a \MQ\ implementation can also be extracted
    from Rhee's algorithm \cite{rhee:fifo},
    from the two algorithms of Lee \cite{hh:qalg},
    as well as from the algorithm of Mellor-Crummey and Scott \cite{mcs:algo}.\footnote
    {The MCS algorithm lacks the bounded exit property, and so 
     the corresponding implementation of \MQ\ is not wait-free due to the presence
     of a busy-wait loop in the access procedure for the \dequeue\ operation.	 
	 In particular, termination of \dequeue\ is only guaranteed if every execution of 
	 \enqueue\ is eventually followed by an execution of \isHead\ by the same process.
	 This condition is certainly satisfied by Algorithm~GQME from Section~\ref{sec_genform}.}

It is interesting to note that the above algorithms are precisely those that
    achieve $O(1)$ RMR complexity in both the CC and DSM models.
Algorithms that are limited to the CC model \cite{graunke:synch, mag:queue}
    tend to have a simpler structure,
    intuitively by taking advantage of the
    fact that any process can locally spin on any variable.
This makes it possible for processes, in particular predecessor-successor pairs,
    to communicate without knowing each other's names.
In particular, in the entry protocol a process can enter the queue
    and signal its predecessor by applying a single atomic operation;
    in contrast, \MQ\ contains distinct operations corresponding to these two tasks.
Also, in the CC model a process in the exit protocol can wake up its successor without
    knowing the successor's identity.
Thus, queue-based local-spin algorithms specific to the CC model operate
    in a mode significantly different from the one captured by the \MQ\ data type.

\section*{Acknowledgments}
I am deeply indebted to Vassos Hadzilacos for his thorough readings of and constructive
feedback on multiple earlier drafts of this paper,
in particular his contributions 
of the informal description of Craig's algorithm at the beginning of Section~\ref{sec_fs},
and extensive help in wording the proofs of correctness in Section{s}~\ref{sec_genform},
    \ref{sec_fi} and \ref{sec_fs}.
I would also like to thank Prasad Jayanti for enlightening discussions about the 
mutual exclusion problem, and in particular for sharing the version
of Craig's mutual exclusion algorithm on which Implementation~MQFS of Section~\ref{sec_fs}
is based.


\newpage
\small
\bibliographystyle{abbrv}
\bibliography{masterbib}

\end{document}